\documentclass[lettersize,journal]{IEEEtran}

\usepackage{amsmath,amsfonts, amssymb, amsthm} 
\usepackage{algorithmicx, algpseudocode, algorithm}
\usepackage{array}
\usepackage[caption=false,font=normalsize,labelfont=sf,textfont=sf]{subfig}
\usepackage{textcomp}
\usepackage{stfloats}
\usepackage{url}
\usepackage{verbatim}
\usepackage{graphicx}
\def\BibTeX{{\rm B\kern-.05em{\sc i\kern-.025em b}\kern-.08em
    T\kern-.1667em\lower.7ex\hbox{E}\kern-.125emX}}
\usepackage{balance}
\usepackage{cite}
\usepackage{adjustbox}

\usepackage{threeparttablex}
\usepackage{tabularx}

\usepackage{mathtools, mathcomp}  
\usepackage{comment}
\usepackage{xcolor}
\usepackage{multirow}
\usepackage{enumerate}
\usepackage{mleftright}
\mleftright

\mathtoolsset{showonlyrefs=true}  

\theoremstyle{definition}
\newtheorem{definition}{Definition}
\newtheorem{remark}{Remark}
\newtheorem{assumption}{Assumption}
\newtheorem{problem}{Problem}

\theoremstyle{plain}
\newtheorem{proposition}{Proposition}
\newtheorem{lemma}{Lemma}
\newtheorem{theorem}{Theorem}

\usepackage[]{hyperref}
\hypersetup{%
 setpagesize=false,%
 bookmarks=true,%
 bookmarksdepth=tocdepth,%
 bookmarksnumbered=true,%
 colorlinks=false,%
 pdftitle={},%
 pdfsubject={},%
 pdfauthor={},%
 pdfkeywords={}
 }

\begin{document}
\title{Simple Controller Design to Achieve Iso-Damping Robustness: Non-Iterative Data-Driven Approach Based on Fractional-Order Reference Model}
\author{Ansei Yonezawa,~\IEEEmembership{Member,~IEEE,} Heisei Yonezawa,~\IEEEmembership{Member,~IEEE,} Shuichi Yahagi,~\IEEEmembership{Member,~IEEE,} \\ and Itsuro Kajiwara

\thanks{\textcolor{red}{This is the accepted version of the paper published in IEEE Transactions on Systems, Man, and Cybernetics: Systems, doi: \href{https://ieeexplore.ieee.org/document/11153074}{10.1109/TSMC.2025.3599208}.}}
\thanks{This work was supported in part by JSPS KAKENHI Grant Number JP23K19084, and in part by the funding provided by ISUZU Advanced Engineering Center. \emph{(Corresponding author: Ansei Yonezawa.)}}
\thanks{Ansei Yonezawa is with the Department of Mechanical Engineering, Kyushu University, Fukuoka, Japan (e-mail: \texttt{ayonezawa[at]mech.kyushu-u.ac.jp}).}
\thanks{Heisei Yonezawa and Itsuro Kajiwara are with the Division of Mechanical and Aerospace Engineering, Hokkaido University, Hokkaido, Japan.}
\thanks{Shuichi Yahagi was with the 6th Research Department, ISUZU Advanced Engineering Center Ltd., Kanagawa, Japan. He is now with the Department of Mechanical Engineering, Tokyo City University, Tokyo, Japan.}
}

\markboth{Journal of \LaTeX\ Class Files}%
{Yonezawa \MakeLowercase{\textit{et al.}}}
\maketitle

\begin{abstract}
  This study proposes a simple controller design approach to achieve a class of robustness, the so-called \emph{iso-damping} property. The proposed approach can be executed using only one-shot input/output data. An accurate mathematical model of a controlled plant is not required. The model-reference control problem is defined to achieve the desired closed-loop specifications, including the iso-damping, and the reference model is designed on the basis of fractional-order calculus. The optimization problem for the model-reference control is formulated using the one-shot input/output data while considering the bounded-input bounded-output (BIBO) stability from a bounded reference input to a bounded output. The iso-damping robust controller is obtained by solving the optimization problem. The representative advantages of the proposed approach over the conventional methods are the simplicity, practicality, and reliability from the viewpoint of the unnecessity of the plant model and explicit consideration of the BIBO stability from a bounded reference input to a bounded output. Numerical and experimental studies demonstrate the validity of the proposed approach.
\end{abstract}


\section{Introduction}\label{S_Introduction}

\subsection{Motivation}\label{S_Motivation}
\IEEEPARstart{R}{obustness} is a fundamental requirement for control systems, as the controlled plant has various uncertainties and characteristic changes. Numerous studies have been conducted to achieve robust control \cite{Safonov2012}. A conventional approach to ensure robustness is 1) modeling the nominal plant and assuming the range of the uncertainty and 2) designing the controller based on the model-based control theory to work well in the pre-specified uncertainty range. However, this approach fails if the plant model including uncertainty is not appropriate \cite{Safonov2018}; modeling the nominal plant and the uncertainty relies on the expert's knowledge, increasing the development cost. Therefore, establishing a practical, simple, and reliable framework for robust controller design is an important challenge.

Among the various uncertainties, gain variation is often found in practical control systems (e.g., load changes in robotic systems \cite{MDMOB2023}). Gain variation adversely affects the control performance and even causes instability. Thus, the robustness to plant gain variation is an indispensable requirement in practical applications. Recently, the \textit{iso-damping} property \cite{CM2005} has garnered a great deal of attention. The iso-damping property is a special class of robustness: the invariance of the overshoot of the setpoint response under plant gain variation. Hence, the iso-damping property is highly desirable for practical control systems, enhancing the reliability and safety.

\begin{remark} \label{Remark_Flat-phase}
    If the Bode phase plot of the open-loop transfer function is locally flat around the gain crossover frequency, such a special characteristic is called the \emph{flat-phase} property \cite{CM2005}. A feedback control system with the flat-phase property achieves the iso-damping robustness \cite{CM2005}. 
\end{remark}

\subsection{Related work}\label{S_Related_Works}
Control systems involving a fractional-order calculus (FC) has generated a great deal of attention: fractional-order (FO) system theory (e.g., stabilization of FO systems \cite{NAJPAA2023, WZ2022, YMZV2024, FLZX2024, LD2024}, stability tests \cite{ZLXJ2022, Yumuk2024, WC2022, MF2022}, the generalized KYP lemma for singular FO systems \cite{LWCW2021}, the FC-based gradient algorithm \cite{WCZC2023}). FC is a mathematical framework generalizing traditional integer-order (IO) calculus, which considers non-integer order differentiations and integrations. Focusing on its rich expression ability and generality, Magin \emph{et al.} have discussed the benefit of exploiting the idea of FC in the cybernetics field \cite{MVP2018}. In particular, FO control is a powerful and effective approach to achieving good control performance and robustness \cite{TAYGHPAC2021}. 
The advantages of FO control have been demonstrated in various industrial applications (e.g., medical devices \cite{Dudhe2025}, thermal power systems \cite{Kumar2025}), highlighting its practical value.
The Bode's ideal transfer function (BITF) is an FO transfer function with the ideal flat-phase specification; it is often utilized for controller design to achieve the iso-damping robustness \cite{BMF2004}. FC-based robust control is one important direction of the exploitation of FC in the cybernetics field.

Various approaches have been proposed to achieve iso-damping on the basis of FO control. These methods aim to tune some design parameters of the fixed structure controller, such as the FO proportional-integral-derivative (PID) controller, in orders to realize the flat-phase property. The analytical conditions for the flat-phase have been derived for a special class of plants and controllers \cite{YGE2019, CL2022, CL2023}. The graphical approach (e.g., plotting the feasible parameter region or achievable performance specification) is an effective way to determine controller parameters satisfying the analytical flat-phase condition \cite{MMNB2020, CLPC2021, WVLCL2024}. These approaches have been established by the frequency domain analyses. It should be noted that both the analytical and graphical approaches in the frequency-domain require accurate mathematical models and parameters of the actual controlled plant. However, the necessity of the plant model and parameters imposes heavy burdens on designers; modeling errors can lead to unexpected deterioration of control performance. Moreover, in most cases, the analytical and graphical policies are derived for very simple plants such as the first-order plus time delay system. Thus, not only the plant model but also its bold approximation is necessary to employ such a method. Controller design based on the approximated model often causes performance degradation due to the gap between the actual plant and the reduced one for controller design.

Numerical optimization-based approaches have been examined for a wide variety of plants in the context of parameter tuning of FO control systems \cite{MMMB2015, PPTJ2024}. As for realizing the flat-phase specification, a novel metaheuristic algorithm has been developed for FO-PID controller tuning \cite{IE2023}. The process of the conventional optimization-based controller tuning can be summarized as follows: (1) perform closed-loop control tests; (2) assess the performance criteria using the results from step (1); (3) determine the next candidate controller based on the evaluation results and the optimization approach; and (4) iterate steps (1)--(3). Nevertheless, this procedure implies that the conventional optimization-based tuning requires huge amount of closed-loop control tests. Conducting control experiments in a real system is time-consuming; simulation-based closed-loop tests require an accurate plant model and parameters for evaluating the objective function.  

Recently, controller design techniques based on the data rather than the mathematical model of the plant have attracted a great deal of attention: direct data-driven control (D3C) \cite{MHD2023, LDWZ2024, ZGGGZCC2023}. Exploiting the D3C approach to design an FO control system can overcome the drawbacks of the abovementioned challenges, since these problems are mainly raised by the needs for accurate plant models and parameters. Several studies have employed frequency response data for designing FO controllers \cite{ST2017, ST2020}. However, obtaining the frequency response data is cumbersome and often requires a special experimental setup.

Model-reference (MR) D3C is a simple and practical framework for achieving the desired closed-loop characteristics while avoiding burdensome frequency response computation. Various MR-D3C techniques have been examined for IO control systems: iterative feedback tuning (IFT) \cite{Hjalmarsson2002}, virtual reference feedback tuning (VRFT) \cite{DF2023, FCCS2019}, correlation-based tuning \cite{vHKB2011}, fictitious reference iterative tuning (FRIT) \cite{Kaneko2013}, to name a few. Many studies have reported the effectiveness of MR-D3C in the IO control framework \cite{YK2022, YK2022_LCSS, WZZSSHF2024, RPPMH2023}. For example, the VRFT approach has been employed for the vehicle yaw rate control \cite{YK2024, YS2023}.

On the other hand, to the best of our knowledge, few studies have explored an MR-D3C approach for FO control. The IFT technique has been employed to tune the FO-PID controller for a hydraulic actuator \cite{MSK2019}. However, the tuning scheme presented in \cite{MSK2019} requires multiple experiments, imposing heavy burdens on designers. The FO-PID controller has been tuned on the basis of the VRFT \cite{AAC2023, AMAMMB2021, XTSZG2018}. Nevertheless, the conventional VRFT does not consider closed-loop stability. 
One study \cite{Xie2019} has proposed a VRFT-based FO controller tuning method with stability considerations; however, the stability verification requires the frequency response data of the controlled plant, which increases the user's burden.
Although one study has tuned the FO-PID controller based on the MR-D3C approach, it does not consider the iso-damping robustness \cite{YYYK2024}. Consequently, to the best of our knowledge, no studies have proposed simple and practical FO MR-D3C techniques especially for achieving the iso-damping property.

\subsection{Contribution and novelty}\label{S_Contribution}
\begin{table*}[]    
  \centering
  \begin{threeparttable}[h]
    \caption{\protect Comparison between the proposed approach and the conventional approaches for iso-damping robustness based on FO system.} 
    \label{Table_Approaches} 
    \begin{tabular}{ll|l|l}
        \hline
        \multicolumn{2}{l|}{\begin{tabular}{l}
    Approach
\end{tabular}} & \begin{tabular}{l}
    Advantage
\end{tabular} & \begin{tabular}{l}
    Challenge
\end{tabular} \\ \hline
        \multicolumn{1}{l|}{\multirow{3}{*}{\begin{tabular}{l}
Conventional
\end{tabular}}} & \begin{tabular}{l}
    Model-based synthesis \\
    in frequency-domain
\end{tabular} & \begin{tabular}{m{5cm}}
    \begin{itemize}
            \item Computationally inexpensive
    \end{itemize}
\end{tabular} & \begin{tabular}{m{5cm}}
    \begin{itemize}
            \item Necessity of plant model
            \item Applicable to simple plants and controllers only
    \end{itemize}
\end{tabular} \\ \cline{2-4} 
        \multicolumn{1}{l|}{} & \begin{tabular}{l}
    Frequency response \\ data-based tuning
\end{tabular} & \begin{tabular}{m{5cm}}
    \begin{itemize}
            \item Free from the necessity of plant model
    \end{itemize}
\end{tabular} & \begin{tabular}{m{5cm}}
    \begin{itemize}
            \item Need for special experimental setup to obtain frequency response data 
    \end{itemize}
\end{tabular} \\ \cline{2-4} 
        \multicolumn{1}{l|}{} & \begin{tabular}{l}
    Minimization of time- \\ domain performance \\ index
\end{tabular} & \begin{tabular}{m{5cm}}
    \begin{itemize}
            \item Broad applicability
            \item Free from the necessity of plant model \tnote{$\dag$}
            
    \end{itemize}
\end{tabular} & \begin{tabular}{m{5cm}}
    \begin{itemize}
            \item Need for iterative closed-loop control tests
            \item Involving non-convex optimization
    \end{itemize}
\end{tabular} \\ \hline
        \multicolumn{1}{l|}{\begin{tabular}{l}
Proposed
\end{tabular}} & \begin{tabular}{l}
    Data-driven approach \\ using fictitious reference
\end{tabular} & \begin{tabular}{m{5cm}}
    \begin{itemize}
            \item Free from the necessity of plant model
            \item Broad applicability
            \item Easy data collection \tnote{$\ddag$}
    \end{itemize}
\end{tabular} & \begin{tabular}{m{5cm}}
    \begin{itemize}
            \item Involving non-convex optimization
    \end{itemize}
\end{tabular} \\ \hline
    \end{tabular}
  \begin{tablenotes}
  \item[$\dag$] This advantage is realized when the performance index is evaluated based on real-world experiments.
  \item[$\ddag$] The proposed approach requires only a single set of time-series input-output data from the controlled plant.
  \end{tablenotes}
  \end{threeparttable}
\end{table*}


This study presents a novel controller design scheme for a discrete-time single-input single-output (SISO) linear time-invariant (LTI) system. The proposed approach does not require a mathematical model of the controlled plant. The present approach can be executed using only one-shot input/output data. The resultant controller due to the proposed approach can achieve iso-damping: the overshoot amount of the closed-loop system is almost invariant under the plant gain variation. In the present approach, the robust controller design problem is defined as the MR-D3C problem using the reference model a priori chosen by the designer. The reference model is designed on the basis of an FO system, defining the desired characteristics of the closed-loop system such as the flat-phase property. The MR-D3C problem is reformulated as an optimization problem based on a fictitious reference signal (the fictitious reference signal is computed using the one-shot input/output data). The loss function to be minimized reflects the information of the closed-loop poles from a reference input to an output. The solution of the minimization problem yields a controller achieving the desired closed-loop characteristic such as the iso-damping specified by the reference model. The validity of the proposed approach is demonstrated through numerical and experimental studies.

Table \ref{Table_Approaches} highlights the distinctions and advancements of the proposed approach over the conventional approaches for iso-damping robust control using FO systems. Compared with the conventional approaches, the contribution and novelty of this study can be summarized as follows:
\begin{enumerate}[(1)]
    \item (\emph{Simplicity}) The proposed approach does not require the mathematical model of the actual controlled plant, providing freedom from costly and burdensome modeling procedures. Therefore, the present approach is simpler than model-based approaches such as analytical and graphical approaches \cite{YGE2019, CL2022, CL2023, MMNB2020, CLPC2021, WVLCL2024} and numerical optimization based on the plant model \cite{BMF2004, IE2023, MMMB2015}. Moreover, the proposed approach is based not on frequency response data but on time-series data, which makes the proposed approach more straightforward than the frequency response data-based approaches \cite{ST2020, ST2017}. This is because obtaining frequency response data is more difficult than obtaining time-series data, as it often requires a special experimental setup.    \label{Contribution_Simplicity}
    \item (\emph{Optimality}) The existing analytical and graphical approaches are developed for very simple plants, requiring not only the plant model but also model approximations. The gap between the actual plant and the approximated model design causes the performance degradation. On the other hand, the proposed technique does not require the model reduction since the plant model itself is unnecessary. Therefore, the present approach is expected to provide superior (i.e., \emph{more} optimal) controller than the approximated-model-based techniques. \label{Contribution_Optimality}
    \item (\emph{Reduced burdens on designers}) The appropriate controller is automatically obtained via solving the optimization problem in the proposed approach. The manual trial-and-error requiring the expert's skill is unnecessary. Moreover, the proposed tuning scheme is executed on the basis of only one-shot input/output data, unlike the traditional FO MR-D3C technique requiring multiple closed-loop tests such as the IFT-based approach \cite{MSK2019}. Thus, the proposed approach significantly reduces the burdens on the designer.    \label{Contribution_Burdens}
    \item (\emph{Reliability}) Unlike the conventional FO MR-D3C techniques based on the VRFT \cite{AAC2023, AMAMMB2021, XTSZG2018}, information regarding the closed-loop pole is explicitly reflected in the proposed optimization problem for the controller design. This feature allows the present technique to explicitly consider the bounded-input bounded-output (BIBO) stability of the closed-loop system from a bounded reference input to a bounded output. Therefore, the proposed design approach yields a reliable controller.    \label{Contribution_Reliability}
    \item (\emph{Generality}) Most analytical approaches and frequency response data-based techniques are developed for a special class of plants and controllers, as discussed above. On the other hand, the proposed approach is applicable to a wide variety of discrete-time SISO LTI plants and controllers.    \label{Contribution_Generality}
\end{enumerate}

Owing to contributions (1)--(5), we can easily incorporate useful and important robustness, i.e., the iso-damping, into various control systems. This study contributes to improve the safety, reliability, and performance of a wide variety of control systems in practice. Moreover, this study has the importance from the viewpoint of the incorporation of FC into the cybernetics field. Specifically, this study is a pioneering work of the exploration the interdisciplinary area between data-driven control and the FO system theory.

\subsection{Structure of this paper}\label{S_Structure}
This paper is organized as follows. Section \ref{S_Preliminaries} summarizes some preliminaries and states the problem addressed in this study. Section \ref{S_Proposed_approach} describes the proposed controller design approach. We show the numerical and experimental studies to demonstrate the validity of the proposed approach in Section \ref{S_Examples}. The discussion is given in Section \ref{S_Discussion}. Section \ref{S_Conclusion} concludes the paper.

\section{Preliminaries}\label{S_Preliminaries}
The symbols $\mathbb{R}$ and $\mathbb{R}_{+}$ denote the sets of real numbers and strictly positive real numbers, respectively. The set of the real-valued matrices with the dimension of $m \times n$ is represented as $\mathbb{R}^{m \times n}$ (we write $\mathbb{R}^{n \times 1}$ as $\mathbb{R}^{n}$). For a vector $v \in \mathbb{R}^{n}$, 
$\mathrm{Tpl}\left(\cdot \right) \colon \mathbb{R} ^{n} \mapsto \mathbb{R} ^{n\times n}$
is defined as
\begin{equation} \label{Eq_Tpl}
    \mathrm{Tpl}(\it{v}) \coloneqq
\begin{bmatrix}
    v_{1} & 0 & 0 & \ddots & 0 \\
    \vdots & v_{1} & 0 & \ddots & 0 \\
    v_{n-2} & \ddots & v_{1} & 0 & \vdots \\
    v_{n-1} & v_{n-2} & \ddots & \ddots & 0 \\
    v_{n} & v_{n-1} & v_{n-2} & \hdots & v_{1} \\
\end{bmatrix}
\end{equation}
where $v_{j} \; (j = 1, 2, \ldots n)$ is the $j$-th component of $v$. For an $n$-length sequence $v = \{v_{k}\}_{k=0}^{n-1} \; (1\leq n\leq \infty)$,
$\mathrm{vec}(v) \coloneqq \begin{bmatrix} v_{0} & v_{1} & \hdots & v_{n-1} \end{bmatrix}^{\top}$
and
$\mathrm{Tpl}\circ \mathrm{vec} \coloneqq \mathrm{Tpl}\left(\mathrm{vec}(\cdot)\right)$.
For a matrix $M \in \mathbb{R}^{n \times n}$, $\Lambda \left(M\right)$ denotes the set of all eigenvalues of $M$ including their algebraic multiplicity. The spectral radius of $M$ is denoted as $\rho\left(M\right)$, i.e.,
$\rho (M) = \max_{\lambda \in \Lambda (M)} \left\lvert \lambda \right\rvert $.

The Laplace variable and the $Z$-variable are represented as $s$ and $z$, respectively. If a transfer function $G\left(\cdot\right)$ has the tunable parameter $\theta \in \mathbb{R}^{n}$, we sometimes denote the transfer function as $G\left(\cdot; \theta \right)$. For a continuous-time transfer function $G \left(s \right)$, $\mathrm{c2d}\left(G\left(s\right)\right)$ represents its discretization. We write $\mathrm{f2i}\left(G_{FO}\left(s\right)\right)$ as the IO approximation of the FO transfer function $G_{FO}\left(s\right)$. Also, 
$\mathrm{c2d}\circ \mathrm{f2i}\left( \cdot \right) \coloneqq \mathrm{c2d}\left( \mathrm{f2i} \left( \cdot \right) \right)$
represents the transformation from a continuous-time FO transfer function to the corresponding discrete-time IO transfer function.

Let $G \left( z \right)$ be a proper rational discrete-time IO SISO LTI system and let $ \{g_{k}\}_{k=0}^{\infty} $ be the impulse response sequence of $G \left( z \right)$.  Then, the output $y = \{y_{k}\}_{k=0}^{\infty}$ of $G \left(z\right)$ to the input $u = \{u_{k}\}_{k=0}^{\infty}$ is $y_{t} = g_{t}\ast u_{t} \; (t=0, 1, 2,\ldots)$. Here, the symbol $\ast$ denotes the convolution, i.e., 
$g_{t}\ast u_{t} \coloneqq \sum_{\tau=0}^{t}g_{\tau}u_{t-\tau} = \sum_{\tau=0}^{t}g_{t-\tau}u_{\tau}$ \cite{FV2012}.
We sometimes denote the input/output relation as $y=G(z)u$. 

The $p$-norm $\left\lVert v \right\rVert _{p}$ of $v \in \mathbb{R}^{n}$ is defined as 
$\left\lVert v \right\rVert _{p} \coloneqq \left(\sum_{k = 1}^{n} \left\lvert v_{j} \right\rvert ^{p} \right)^\frac{1}{p}$ 
for $1\leq p < \infty$ and 
$\left\lVert v \right\rVert _{\infty} \coloneqq \sup_{j}\left\lvert v_{j} \right\rvert$. 
Similarly, the $l_{p}$-norm $\left\lVert u \right\rVert _{p}$ of a real-valued signal $u = \{u_{k}\}_{k=0}^{\infty}$ is defined as 
$\left\lVert u \right\rVert _{p} \coloneqq \left(\sum_{k = 0}^{\infty} \left\lvert u_{j} \right\rvert ^{p} \right)^\frac{1}{p}$ 
for $1\leq p < \infty$ and 
$\left\lVert u \right\rVert _{\infty} \coloneqq \sup_{j}\left\lvert u_{j} \right\rvert$. 

\subsection{Flat-phase property and iso-damping}\label{S_Flat_phase}
\begin{figure}[!t]
    \centering
    \includegraphics[width=6.8cm]{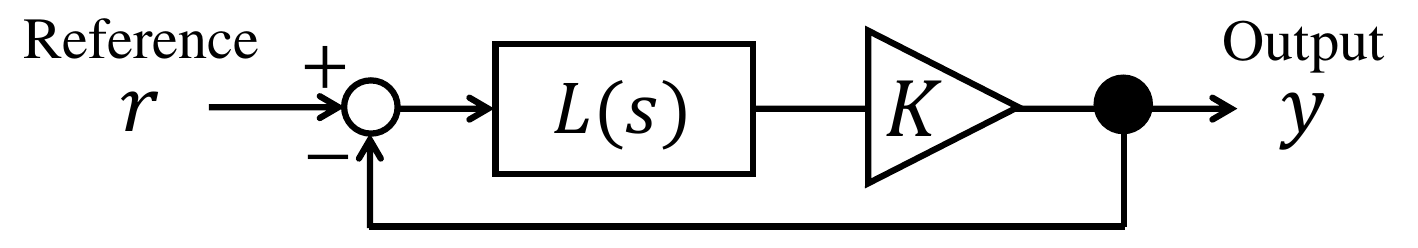}
    \caption{\protect{Closed-loop system.}}
    \label{Fig_Closed-loop}
\end{figure}

Consider the feedback control system shown in Fig. \ref{Fig_Closed-loop}, where $L\left(s\right)$ is a continuous-time SISO LTI transfer function and $K$ is the static gain. Let $\omega_{c} \in \mathbb{R}_{+}$ be such that $\left\lvert L \left(j \omega_{c} \right) \right\rvert =1 $. The flat-phase property can be characterized as
\begin{equation} \label{Eq_Flat-phase}
    \left.\frac{d}{ds} \left(\angle L(s)\right)\right|_{s=j\omega_{c} } = 0
\end{equation}
where $\angle L\left(s\right)$ represents the argument of $L \left(s\right)$ \cite{CM2005}. If $L\left(s\right)$ in Fig. \ref{Fig_Closed-loop} has the flat-phase property, the closed-loop system shown in Fig. \ref{Fig_Closed-loop} has a special robustness: the overshoot amount of the output $y$ to the setpoint reference input $r$ is invariant under the variation of $K$, namely, the iso-damping \cite{CM2005}. The iso-damping property is practically useful, as the plant gain variation often occurs in many control systems. 

\begin{remark} \label{Remark_BITF}
    A typical choice of $L\left(s\right)$ realizing the flat-phase property is the BITF $L_{Bode}\left(s\right)$:
        \begin{equation} \label{Eq_BITF}
        L_{Bode}\left(s\right) = \left(\frac{\omega _{c}}{s}\right)^{\gamma}
    \end{equation}
    where $\gamma \in \mathbb{R}$. Note that the BITF is an FO transfer function because the FO integrator/differentiator becomes $s^{\alpha} \; (\alpha \in \mathbb{R})$ in the Laplace domain ($\alpha < 0$ for integrator, $\alpha > 0$ for differentiator). This fact clearly shows that the FO control framework is effective to realize the iso-damping property, owing to its high design flexibility. The design policy for the BITF has been discussed in \cite{BMF2004}. 
    Further details of the Laplace transforms of FO derivatives and integrals can be found, for example, in Section 2.8 of \cite{Podlubny1998}.  
\end{remark}

\subsection{BIBO stability of proper rational discrete-time IO SISO LTI system}\label{S_BIBO}
In this study, we employ the standard definition of the BIBO stability \cite{FV2012}. Let $G \left(z\right)$ be a proper rational discrete-time IO SISO LTI transfer function. Then, the BIBO stability of $G \left(z\right)$ is defined as follows.

\begin{definition} \label{Def_BIBO}
    Let $\left\{g_{k}\right\}_{k=0}^{\infty}$ be the impulse response sequence of $G \left(z\right)$ and let $\left\{u_{k}\right\}_{k=0}^{\infty}$ be an arbitrary bounded input sequence, i.e., there exists $0 < \alpha_{u} < \infty$ such that 
    $\left\lvert u_{k} \right\rvert  < \alpha_{u} \; (k=0,1,2,\ldots)$. 
    Then, $G \left(z\right)$ is said to be BIBO stable if there exists $0 < \alpha_{y} < \infty$ satisfying $\left\lvert g_{k}\ast u_{k} \right\rvert  < \alpha_{y} \; (k=0,1,2,\ldots)$. 
\end{definition}

The following facts are standard.

\begin{proposition} \label{Prop_BIBO_iff}
    The transfer function $G\left(z\right)$ is BIBO stable if and only if one of the following conditions is satisfied: 
    \begin{enumerate}[(C1)]
    \item Its impulse response sequence $g = \left\{g_{k}\right\}_{k=0}^{\infty}$ is absolutely summable, i.e., $ \left\lVert g \right\rVert _{1} < \infty $;
    \item All poles of $G\left(z\right)$ exist inside the unit circle.
    \end{enumerate}
\end{proposition}
\emph{Proof:} See Section 4.3.2 in \cite{FV2012}. \qed

\subsection{Overview of the controller design problem in this study}\label{S_Overview_of_problem}
\begin{figure}
    \begin{center}
    \includegraphics[width=6.7cm]{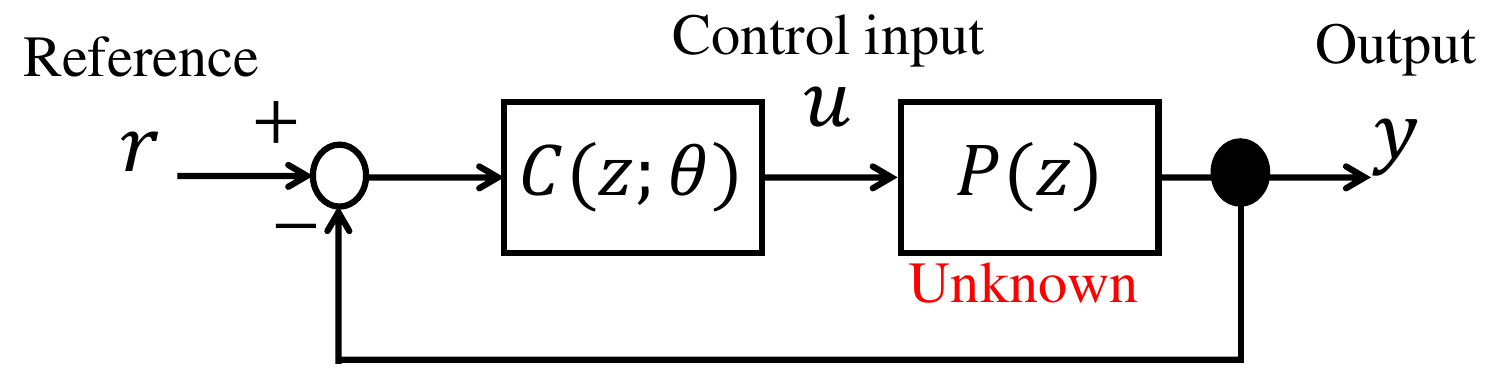}
    \caption{\protect{Closed-loop system with unknown plant $P \left(z\right)$ and tunable controller $C \left(z; \theta\right)$.}}
    \label{Fig_Closed-loop_w_P_C}
    \end{center}
\end{figure}

Fig. \ref{Fig_Closed-loop_w_P_C} describes the control system considered in this study. In Fig. \ref{Fig_Closed-loop_w_P_C}, the plant to be controlled has transfer function $P\left(z\right)$; $C\left(z; \theta\right)$ is the controller having the tunable parameter $\theta \in \Theta \subset \mathbb{R} ^ {n}$, where $\Theta$ is the search range for $\theta$. Here, $P\left(z\right)$ is a proper rational discrete-time IO SISO LTI system; $C\left(z; \theta\right)$ is a proper rational discrete-time IO SISO LTI system for any $\theta$. The class of the controller is a priori given by the designer. The setpoint reference signal, the control input, and the output of the actual system are denoted as $r$, $u$, and $y$, respectively. The reference signal $r$ is assumed to be bounded. The closed-loop transfer function from $r$ to $y$ is denoted as
$T\left( z; \theta \right) = P\left(z\right)C\left(z; \theta\right)\left\{1+P\left(z\right)C\left(z; \theta\right)\right\}^{-1}$.

Here, the input data $u_{[0:N]}^{D} = \left\{ u_{k}^{D} \right\}_{k=0}^{N}$ and the output data $y_{[0:N]}^{D} = \left\{ y_{k}^{D} \right\}_{k=0}^{N}$ of the controlled plant $P \left( z \right)$ are collected through an open-loop or closed-loop experiment. In this study, we address the problem of finding $\theta$ such that $T\left(z; \theta\right)$ satisfies the designer-specified design specification including the iso-damping property, where $P\left(z\right)$ is unknown. Throughout this study, the input/output data is assumed to be noiseless. Under noisy conditions, the total variation denoising \cite{YK2021_MC} is a practical choice to deal with the measurement noise.

\section{Proposed approach}\label{S_Proposed_approach}
This section presents the controller tuning problem described in Section \ref{S_Overview_of_problem} in order to achieve the iso-damping robust control.

\subsection{Robust controller design via MR-D3C framework}\label{S_MR-D3C}
\begin{figure}
    \begin{center}
    \includegraphics[width=8.5cm]{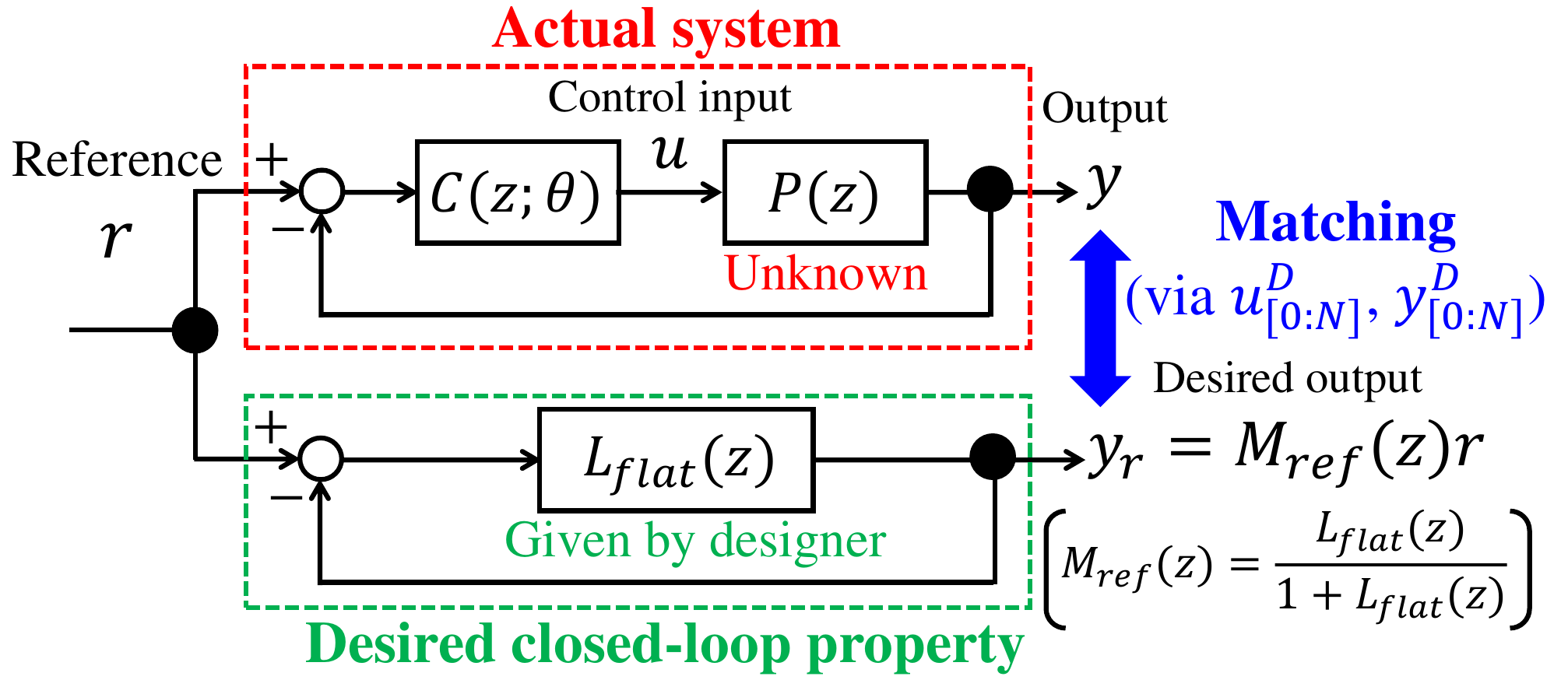}
    \caption{\protect{MR-D3C approach to design a controller to achieve the iso-damping robustness.}}
    \label{Fig_MR-D3C}
    \end{center}
\end{figure}

Fig. \ref{Fig_MR-D3C} sketches the proposed controller design approach based on the MR-D3C framework. Let $L_{flat}\left(s\right)$ be the continuous-time FO SISO LTI transfer function satisfying \eqref{Eq_Flat-phase}, and $L_{flat}\left(z\right) \coloneq \mathrm{c2d}\circ \mathrm{f2i}\left(L_{flat}\left(s\right) \right)$.  In Fig. \ref{Fig_MR-D3C}, $M_{ref} \left( z \right) = L_{flat} \left( z \right) \left\{1 + L_{flat}\left( z \right) \right\} ^ {-1}$ is the FC-based reference model specifying the desired closed-loop characteristics. Note that $M_{ref} \left( z \right)$ has the iso-damping robustness owing to the specification of $L_{flat}\left(z\right)$. The designer a priori determines the specific choice of $L_{flat}\left(s\right)$. The meaning of $P\left(z\right)$, $C\left(z; \theta \right)$, $r$, $u$, and $y$ are the same as those in Fig \ref{Fig_Closed-loop_w_P_C}. The following assumption can be easily satisfied by suitably designing $L_{flat}\left(z\right)$.

\begin{assumption} \label{Assumpion_L-flat}
    $L_{flat}\left(z\right)$ is proper and rational. $M_{ref}\left(z\right)$ has all poles inside the unit circle, implying $M_{ref}\left(z\right)$ is BIBO stable. 
\end{assumption}

Then, the controller design problem in Section \ref{S_Overview_of_problem} can be formally formulated as follows:

\begin{problem}[MR-D3C for iso-damping] \label{Problem_MR-D3C}
    Find the optimal parameter $\theta^{\ast } \in \mathbb{R} ^{n}$ of the controller on the basis of $u_{[0:N]}^{D}$ and $y_{[0:N]}^{D}$, such that $T \left( z; \theta^{\ast} \right)r_{[0:N]}^{D}$ is as close to $M_{ref}\left( z \right) r_{[0:N]}^{D}$ as possible in the mean square error sense. 
    Here, $r_{[0:N]}^{D} = \left\{ r_{k}^{D} \right\}_{k=0}^{N} \, \mleft( r_{0}^{D}\neq 0 \mright)$ is the reference input specified by the designer.
\end{problem}

\begin{remark} \label{Remark_FOC} 
    As $L_{flat}\left(z\right)$ is constructed based on the FO transfer function, it is preferable that the class of the controller is also chosen as the FC-based controller such as
    $C \left( z; \theta \right) = \mathrm{c2d}\circ \mathrm{f2i}\left( C_{FOPID} \left( s; \theta \right) \right)$, 
    where $C_{FOPID} \left( s; \theta \right)$ is the standard FO-PID controller with the tunable parameter $\theta$. This fact is demonstrated in Example 1 in Section \ref{S_Examples}. 
\end{remark}

\subsection{Fictitious reference signal}\label{S_Fictitious_reference_singal}
The proposed approach addresses Problem \ref{Problem_MR-D3C} using the fictitious reference signal \cite{Kaneko2013, ST1997}. The fictitious reference signal 
$\tilde{r}_{[0:N]} \left( \theta \right) = \left\{ \tilde{r}_{k} \left(\theta\right) \right\}_{k=0}^{N}$ 
is defined as 
\begin{equation}\label{Eq_r_tilde}
    \tilde{r}_{[0:N]} \left( \theta \right) = \left( C \left( z; \theta \right) \right)^{-1} u_{[0:N]}^{D} + y_{[0:N]}^{D}.
\end{equation}
The fictitious reference signal has a unique property as follows:
\begin{lemma} \label{Lemma_Ficref}
    For any $\theta$ such that $1 + P \left( z \right)C \left( z; \theta \right) \neq 0$, $T \left( z; \theta \right) \tilde{r}_{[0:N]} \left( \theta \right) = y_{[0:N]}^{D}$.
\end{lemma}
\emph{Proof:} The statement follows from straightforward calculations \cite{Kaneko2013}. In particular, 
\begin{align}\label{Eq_Proof_Lemma_Ficref}
    & T\left(z; \theta \right) \tilde{r}_{[0:N]} \left( \theta \right) \\
&= \frac{P\left(z\right) C \left( z; \theta \right)}{1+P\left(z\right) C \left( z; \theta \right)}
\left\{ \left( C \left( z; \theta \right) \right)^{-1} u_{[0:N]}^{D} + y_{[0:N]}^{D} \right\} \\
&= \frac{1}{1+P\left(z\right) C \left( z; \theta \right)} P\left(z\right) u_{[0:N]}^{D}
   + \frac{P\left(z\right) C \left( z; \theta \right)}{1+P\left(z\right) C \left( z; \theta \right)} y_{[0:N]}^{D} \\
&= \frac{1}{1+P\left(z\right) C \left( z; \theta \right)}
   \left\{ y_{[0:N]}^{D} + P\left(z\right) C \left( z; \theta \right) y_{[0:N]}^{D} \right\} \\
&= y_{[0:N]}^{D}
\end{align}     
which is the desired outcome. \qed

Hereafter, we assume that $\tilde{r}_{0} \left( \theta \right) \neq 0$ for any $\theta$.
\begin{remark} \label{Remark_FRIT}
    One study has proposed a MR-D3C technique based on the fictitious reference signal: FRIT \cite{Kaneko2013}. However, the controller provided by FRIT may destabilize the closed-loop system as pointed out in \cite{YK2021_MC, SYK2022}. To overcome this difficulty, several studies have examined to modify FRIT for IO control \cite{YK2021_MC, SYK2022}. 
\end{remark}

\subsection{Proposed controller design approach}\label{S_Proposed_controller_design_approach}
In this study, we propose a novel approach for achieving iso-damping robust control in the MR-D3C framework. Specifically, to solve Problem \ref{Problem_MR-D3C}, we propose a new MR-D3C technique: instability detecting FRIT for the iso-damping robust control (Iso-IDFRIT). 
\begin{theorem}[Iso-IDFRIT] \label{Theorem_iso-IDFRIT}
    Let $M_{ref} \left(z\right)$ be the reference model constructed based on the FO open-loop transfer function with the flat-phase property, as stated in Section \ref{S_MR-D3C}. Let $m^{ref} = \left\{ m_{k}^{ref} \right\}_{k=0}^{\infty}$ be the impulse response sequence of $M_{ref} \left(z\right)$ and $m_{[0:N]}^{ref} \coloneqq \left\{ m_{k}^{ref} \right\}_{k=0}^{N}$. Then, the solution of Problem \ref{Problem_MR-D3C} is given by
    \begin{equation} \label{Eq_minimize_J}
        \theta^{\ast} = \underset{\theta}{\arg\min} \: J \left( \theta \right) 
    \end{equation}
    \begin{multline} \label{Eq_Loss_iso-IDFRIT} 
        J \left( \theta \right) 
\\= \left\lVert R^{D} \left( \tilde{R} \left( \theta \right) \right)^{-1} \mathrm{vec} \left( y_{[0:N]}^{D} \right) 
\right.
\left.
\vphantom{\left( \tilde{R} \left( \theta \right) \right)^{-1}}
- R^{D} \mathrm{vec} \left( m_{[0:N]}^{ref} \right) \right\rVert _{2}^{2} 
    \end{multline}
    where $R^{D} = \mathrm{Tpl} \circ \mathrm{vec} \left( r_{[0:N]}^{D} \right)$ and 
    $\tilde{R} \left( \theta \right) = \mathrm{Tpl} \circ \mathrm{vec} \left( \tilde{r}_{[0:N]} \left( \theta \right)  \right)$. 
\end{theorem}
\emph{Proof:} Let $t \left( \theta \right) = \left\{ t_{k} \left( \theta \right) \right\}_{k=0}^{\infty}$ be the impulse response sequence of $T \left(z; \theta \right)$. Due to Lemma \ref{Lemma_Ficref}, 
\begin{equation} \label{Eq_Vec_ficref}
    \mathrm{vec} \left( y_{[0:N]}^{D} \right) = \tilde{R} \left( \theta \right) \mathrm{vec} \left( t_{[0:N]} \left( \theta \right) \right)
\end{equation}
where $t_{[0:N]} \left( \theta \right) \coloneqq \left\{ t_{k} \left( \theta \right) \right\}_{k=0}^{N}$. Since $\tilde{r}_{0} \left( \theta \right) \neq 0$, the impulse response of $T \left(z; \theta \right)$, which is the closed-loop transfer function from the reference input to the output when the controller parameter is $\theta$, from time $0$ to $N$ is restored as 
\begin{equation} \label{Eq_Deconvolution}
    \mathrm{vec} \left( t_{[0:N]} \left( \theta \right) \right) 
=
\left( \tilde{R} \left( \theta \right) \right) ^{-1} \mathrm{vec} \left( y_{[0:N]}^{D} \right).
\end{equation}
Here, let $y_{[0:N]} \left( \theta \right) = \left\{ y_{k} \left( \theta \right) \right\}_{k=0}^{N}$ denote the output of $T \left(z; \theta \right)$ due to $r_{[0:N]}^{D}$. Using \eqref{Eq_Deconvolution}, $y_{[0:N]} \left( \theta \right)$ can be estimated as
\begin{align} \label{Eq_DD-estimation}  
    \mathrm{vec} \left( y_{[0:N]} \left( \theta \right) \right) &= R^{D} \mathrm{vec} \left( t_{[0:N]} \left( \theta \right) \right) \notag \\   
&= R^{D} \left( \tilde{R} \left( \theta \right) \right)^{-1} \mathrm{vec} \left( y_{[0:N]}^{D} \right).
\end{align}
On the other hand, the output of $M_{ref} \left(z\right)$ driven by $r_{[0:N]}^{D}$ can be computed as 
\begin{equation} \label{Eq_Mref-rd}
    \mathrm{vec} \left( M_{ref}\left(z\right) r_{[0:N]}^{D} \right) 
= 
R^{D} \mathrm{vec} \left( m_{[0:N]}^{ref}  \right)
\end{equation}
which completes the proof. \qed

Note that the minimization problem \eqref{Eq_minimize_J} is formulated in purely data-driven manner, i.e., the plant model $P\left(z\right)$ does not appear in \eqref{Eq_minimize_J}. The desired closed-loop property specified by $M_{ref} \left(z\right)$, which has the iso-damping property, is realized by $C \left( z; \theta^{\ast} \right)$, since $\theta^{\ast}$ achieves the model matching between $M_{ref} \left(z\right)$ and $T \left( z; \theta^{\ast} \right)$ as closely as possible in the predetermined controller class. Therefore, the proposed approach, Iso-IDFRIT, can design the desired controller $C \left( z; \theta^{\ast} \right)$ that achieves the iso-damping robustness without knowing the mathematical model $P\left(z\right)$ of the controlled plant.

\begin{remark} \label{Remark_BIBO_of_T}
    Equation \eqref{Eq_DD-estimation} shows that the loss function $J\left(\theta\right)$ in the minimization problem \eqref{Eq_minimize_J} explicitly reflects the impulse response of $T \left( z; \theta \right)$. If the value of $J\left(\theta\right)$ is reasonably small, we can consider that $T \left( z; \theta \right)$ is BIBO stable. Conversely, the controller making the value of $J\left(\theta\right)$ excessively large should destabilize the closed-loop system. That is, we can evaluate the BIBO stability of the resultant closed-loop system from a bounded reference input to a bounded output \emph{before} implementing the controller. Therefore, iso-IDFRIT can design a reliable and practically feasible controller.
\end{remark}

\subsection{Stability analysis}\label{S_Stability_analysis}
In this section, we analyze the BIBO stability of $T \left( z; \theta^{\ast} \right)$, i.e., the closed-loop system given by the proposed method, in an asymptotic case ($N \to \infty$). Here, we assume the following assumption:
\begin{assumption} \label{Assumption_J_finite}
    The optimization problem \eqref{Eq_minimize_J} is successfully solved for the asymptotic case. That is, $J\left(\theta\right) \to \alpha$ for $N \to \infty$, where $\alpha \in \mathbb{R} _{+}$ is a finite value. 
\end{assumption}

The following lemma is instrumental for the stability analysis:

\begin{lemma} \label{Lemma_BIBO-State-Space}
    Let $G\left(z\right)$ be a proper rational discrete-time IO SISO LTI transfer function. The impulse response sequence of $G\left(z\right)$ is denoted as $g = \left\{ g_{k} \right\}_{k=0}^{\infty}$. Then, $G\left(z\right)$ is BIBO stable if and only if $\left\lVert g \right\rVert _{2} < \infty$. 
\end{lemma}

\emph{Proof:} Let $(A,B,C,D)$ be the the minimal state-space realization of $G\left(z\right)$, i.e., $G(z) = C \left( zI-A \right)^{-1} B + D$ ($I$ is the identity matrix with the compatible dimension). Then, $g_{0} = D$ and $g_{k} = CA^{k-1}B$ for $k\geq 1$. Note that $\Lambda\left(A\right)$ coincides the poles of $G\left(z\right)$ including multiplicity due to the minimality of $(A,B,C,D)$ (Theorem 12.9.16 in \cite{Bernstein2009}). This implies that $G\left(z\right)$ is BIBO stable if and only if $\rho \left( A \right) < 1$ due to Proposition \ref{Prop_BIBO_iff}.

Assume that $G\left(z\right)$ is BIBO stable. Then, due to Proposition \ref{Prop_BIBO_iff}, 
$\left( \left\lVert g \right\rVert _{2}  \right)^{2} < \left\lVert g \right\rVert _{\infty} \left\lVert g \right\rVert _{1} < \infty$. In contrast, suppose that $G\left(z\right)$ is not BIBO stable, i.e., $\rho \left( A \right) \geq 1$. Here, for $k \to \infty$, $\left\lvert g_{k} \right\rvert ^{2} = \left\lvert CA^{k-1}B \right\rvert ^{2} \to 0$ if and only if $A^{k-1} \to 0$. However, $A^{k-1} \to 0$ if and only if $\rho \left( A \right) < 1$ (Theorem 5.6.12 in \cite{HJ2012}), proving that $\left\lVert g \right\rVert _{2}$ is not finite. \qed

Here, the main result of the stability analysis is as follows:

\begin{theorem} \label{Theorem_Stability_analysis}
    Under the Assumptions \ref{Assumpion_L-flat} and \ref{Assumption_J_finite}, $T \left(z; \theta^{\ast}\right)$ is BIBO stable. 
\end{theorem}

\emph{Proof:} Let $\varepsilon _{[0:N]} \left( \theta^{\ast} \right)$ denote the error between the output of the closed-loop system $T \left( z; \theta^{\ast} \right)$ driven by $r_{[0:N]}^{D}$ and the desired closed-loop response due to $r_{[0:N]}^{D}$. That is, 
\begin{equation} \label{Eq_epsilon_def}
    \varepsilon _{[0:N]} \left( \theta^{\ast} \right) \coloneqq T \left( z; \theta^{\ast} \right)r_{[0:N]}^{D}- M_{ref} \left( z \right) r_{[0:N]}^{D}.
\end{equation}    
Then,
\begin{multline} \label{Eq_Vec-epsilon}
    \mathrm{vec} \left( \varepsilon _{[0:N]} \left( \theta^{\ast} \right) \right) 
= 
R^{D} \mathrm{vec} \left( t \left( z; \theta^{\ast} \right) \right)
-
R^{D} \mathrm{vec} \left( m_{[0:N]} ^ {ref} \right).
\end{multline}
That is,
\begin{multline} \label{Eq_Vec-t} 
    \mathrm{vec} \left( t_{[0:N]} \left( \theta^{\ast} \right) \right) 
= 
\left( R^{D} \right) ^{-1} \mathrm{vec} \left( \varepsilon_{[0:N]}  \left(\theta^{\ast} \right) \right)
+
\mathrm{vec} \left( m_{[0:N]} ^ {ref} \right).
\end{multline}
Here, 
\begin{equation}
J \left( \theta ^{\ast} \right) = \left\lVert \varepsilon_{[0:N]} \left(\theta^{\ast} \right) \right\rVert _{2} ^{2} 
= \left\lVert \mathrm{vec} \left( \varepsilon_{[0:N]}  \left(\theta^{\ast} \right) \right) \right\rVert _{2} ^{2}
\end{equation}
due to \eqref{Eq_Deconvolution} and \eqref{Eq_Vec-epsilon}. 
Therefore, 
\begin{equation} \label{Eq_Vec-t_upper} 
    \begin{split}
&\left\lVert 
    \mathrm{vec} \left( t_{[0:N]} \left( \theta^{\ast} \right) \right) 
\right\rVert _{2} \\
&= 
\left\lVert 
    \left( R^{D} \right) ^{-1} \mathrm{vec} \left( \varepsilon_{[0:N]}  \left(\theta^{\ast} \right) \right)
    +
    \mathrm{vec} \left( m_{[0:N]} ^ {ref} \right) 
\right\rVert _{2} \\
&\leq 
\beta _{RD} \sqrt{J\left(\theta^{\ast}\right)}
+
\left\lVert 
    \mathrm{vec} \left( m_{[0:N]} ^ {ref} \right) 
\right\rVert _{2}
\end{split}
\end{equation}
where $\beta_{RD} \in \mathbb{R}_{+}$ denotes the induced 2-norm of $\left(R^{D}\right)^{-1}$. Here, there exists a finite $\beta_{ref} \in \mathbb{R}_{+}$ such that 
$\lim_{N\to\infty}\left\lVert \mathrm{vec} \left( m_{[0:N]}^{ref} \right) \right\rVert _{2} = \left\lVert m^{ref} \right\rVert _{2} < \beta _{ref} $ due to Assumption \ref{Assumpion_L-flat} and Lemma \ref{Lemma_BIBO-State-Space}. Finally,
\begin{equation} \label{Eq_Vec-t_finite} 
    \begin{split}
\lim _{N \to \infty}
\left\lVert 
    \mathrm{vec} \left( t_{[0:N]} \left( \theta^{\ast} \right) \right) 
\right\rVert _{2}
&= 
\left\lVert 
    t \left(\theta^{\ast} \right) 
\right\rVert _{2} \\
&\leq 
\sqrt{\alpha} \beta _{RD}  
+
\beta_{ref}
<
\infty
\end{split}
\end{equation}
which proves the conclusion. \qed

\begin{remark} \label{Remark_Finite-data}
    In Theorem \ref{Theorem_Stability_analysis}, Assumption \ref{Assumption_J_finite} is a sufficient condition for proving the BIBO stability of $T \left(z; \theta^{\ast}\right)$; the BIBO stability may be achieved even if Assumption \ref{Assumption_J_finite} does not hold. 
    Moreover, Theorem \ref{Theorem_Stability_analysis} is the asymptotic result. Analysis in the finite sample case and the derivation of the necessary and sufficient condition of the BIBO stability are left as future work. Nevertheless, the proposed loss function $J\left(\theta\right)$ in \eqref{Eq_Loss_iso-IDFRIT} explicitly reflects the pole information of $T \left(z; \theta^{\ast}\right)$ via its $\left(N+1\right)$-length impulse response $t_{[0:N]} \left( \theta \right)$. This feature allows us to \emph{infer} that the resultant controller $C \left(z; \theta^{\ast}\right)$ stabilizes the closed-loop system from the reference input to the output in the BIBO sense, improving the reliability of the tuning result.
\end{remark}

\subsection{Summary of the proposed controller design}\label{S_Summary_of_iso-IDFRIT}
\begin{figure}
    \begin{center}
    \includegraphics[width=7.2cm]{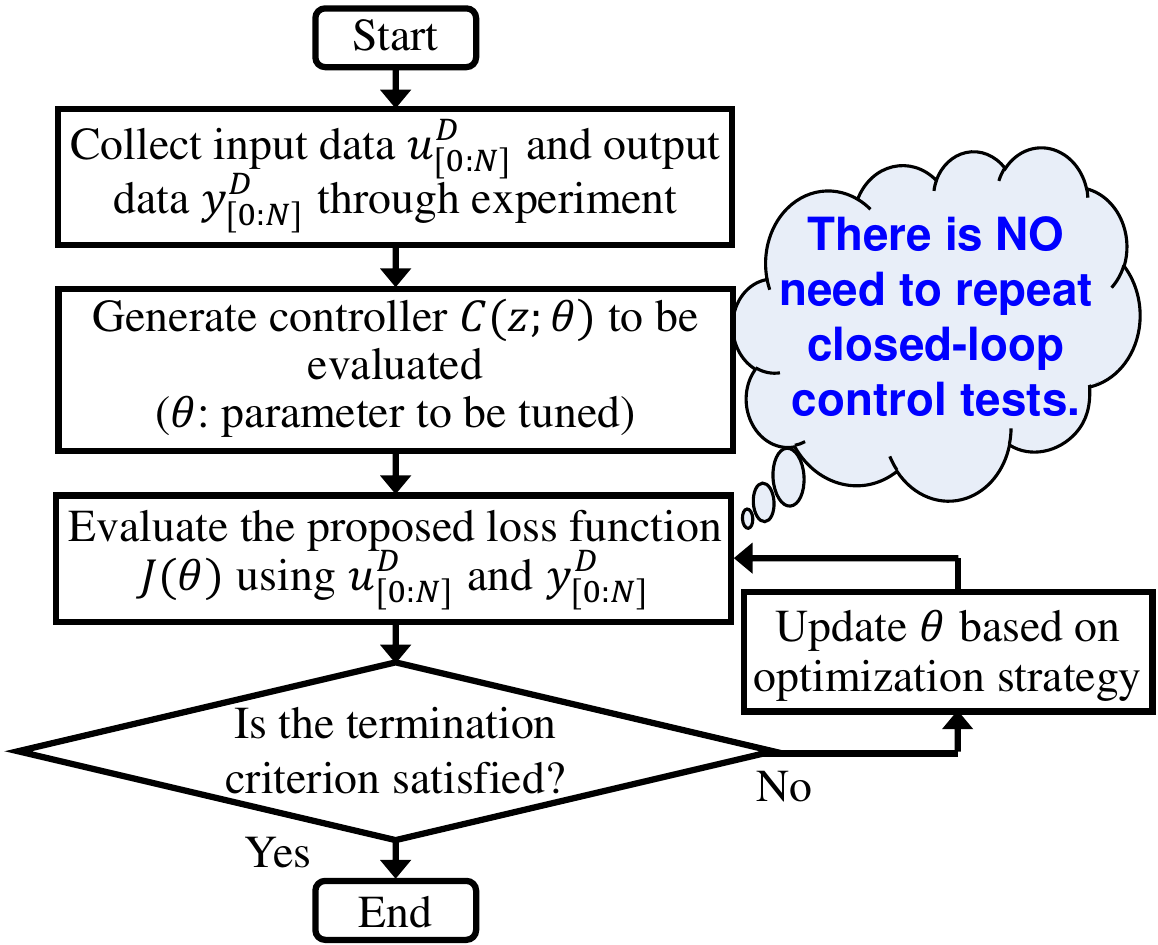}
    \caption{\protect{
            Solution procedure of the optimization problem for the proposed controller design approach. The loss function $J \left( \theta \right)$ shown in \eqref{Eq_Loss_iso-IDFRIT} can be minimized without iteratively conducting closed-loop control tests.
    }}
    \label{Fig_Flowchart}
    \end{center}
\end{figure}

Procedure \ref{Procedure_iso-IDFRIT} summarizes the proposed data-driven design method. In Procedure \ref{Procedure_iso-IDFRIT}, $C_{FO}\left(s; \theta\right)$ is an FO controller with the tunable parameter $\theta$. Procedure \ref{Procedure_iso-IDFRIT} can handle various types of controllers and reference models; the present approach is highly versatile and flexible. Figure \ref{Fig_Flowchart} describes the solution procedure of the optimization problem for the proposed controller design approach. The fictitious reference signal $\tilde{r}_{[0:N]} \left( \theta \right)$, which is used to compute $J\left(\theta\right)$, can be generated straightforwardly by filtering $u_{[0:N]}^{D}$ by $\left( C \left( z; \theta \right) \right)^{-1}$ and adding it to $y_{[0:N]}^{D}$, as shown in \eqref{Eq_r_tilde}.
\begin{algorithm}
    \floatname{algorithm}{Procedure} 
    \begin{enumerate}
        \item Determine the control specifications including the iso-damping robustness. Determine the IO approximation and the discretization methods.
        \item $L_{flat}\left( s \right)$ 
is given as the FO transfer function to to satisfy the specifications, and 
$L_{flat}\left( z \right) = \mathrm{c2d} \circ \mathrm{f2i} \left( L_{flat}\left( s \right) \right)$.
The reference model is given as
$M_{ref}\left(z\right) = L_{flat}\left( z \right) \left\{ 1+ L_{flat}\left( z \right) \right\}$.

        \item Define the class of controller 
$C\left( z; \theta \right)$ 
, e.g., 
$C\left(z; \theta \right) = \mathrm{c2d} \circ \mathrm{f2i} \left( C_{FO} \left(s; \theta\right) \right)$.

        \item Acquire the initial data
$u_{[0:N]}^{D}$ 
and
$y_{[0:N]}^{D}$ 
through an open-loop or closed-loop experiment. 
If these data are corrupted by noise, mitigate the data noise using a denoising technique (e.g., low-pass filtering, the total variation denoising \cite{YK2021_MC}).

        \item Solve the optimization problem \eqref{Eq_minimize_J} for the reference input $r_{[0:N]}^{D}$. 
The solution of \eqref{Eq_minimize_J} is denoted as 
$\theta^{\ast}$.
        \item Return 
$C \left(z; \theta^{\ast}\right)$
as the resultant controller.
    \end{enumerate}
\caption{Proposed approach (Iso-IDFRIT).} 
\label{Procedure_iso-IDFRIT}
\end{algorithm}

\begin{remark} \label{Remark_Frequency}
    We can interpret the loss function $J\left(\theta\right)$ from the viewpoint of the frequency domain. Specifically, since $J \left( \theta \right) = \sum_{n = 0}^{N} \left\lvert \left( t_{k} \left( \theta \right) - m_{k}^{ref} \right) \ast r_{k}^{D} \right\rvert ^{2} $,
    \begin{multline} \label{Eq_Fourier}  
        \lim _{N \to \infty} J \left( \theta \right) 
= \sum_{n = 0}^{\infty} \left\lvert \left( t_{k} \left( \theta \right) - m_{k}^{ref} \right) \ast r_{k}^{D} \right\rvert ^{2}
\\
= 
\frac{1}{2\pi}\int_{-\pi}^{\pi} \left\lvert T\left( e^{j\omega}; \theta \right) - M_{ref} \left( e^{j\omega} \right) \right\rvert \Phi _{rD} \left( e^{j\omega} \right) \,d\omega 
    \end{multline}
    where $\Phi_{rD} \left(e^{j\omega}\right) $ is the spectral density of the reference input $r_{[0:\infty]}^{D}$. Therefore, the minimization of $J\left(\theta\right)$ corresponds to the minimization of the mismatch of the frequency responses $T\left(e^{j\omega}; \theta\right)$ and $M_{ref}\left(e^{j\omega}\right)$ with the frequency weighting function $\Phi_{rD} \left(e^{j\omega}\right) $. 
\end{remark}

\begin{remark} \label{Remark_Reliability}
    In practice, the IO approximation and discretization have to be carried out in order to implement an FO controller \cite{TAYGHPAC2021, CPX2009}. The IO approximation and discretization may affect the closed-loop performance and stability \cite{DATK2020}. Unlike conventional approaches, the proposed design scheme explicitly handles the IO-approximated and discretized controller. Specifically, the proposed approach evaluates the resultant closed-loop performance and stability for the IO-approximated and discretized, i.e., \emph{ready-to-implement}, controller, showing the reliability of the proposed approach.
\end{remark}

\section{Validation}\label{S_Examples}
The effectiveness of the proposed approach is demonstrated in this section. Throughout this study, the Oustaloup recursive filter \cite{OLMN2000, Tepljakov2017} is employed for the IO approximation $\mathrm{f2i}$ for both the reference model and the controller; we use the Tustin method for all $\mathrm{c2d}$ schemes. The design and IO approximation of the FO transfer function are performed via the FOMCON toolbox \cite{TPB2011}. The iso-damping reference model is designed using the BITF shown in \eqref{Eq_BITF}, where the integral order and the gain crossover frequency are denoted as $\gamma$ and $\omega_{c}$ $\mathrm{[rad/s]}$, respectively. Here, $\gamma = 2 \left( 1- \frac{\phi_{m}}{\pi} \right)$, where $\phi_{m}$ $\mathrm{[rad/s]}$ represents the phase margin \cite{BMF2004}. That is, $M_{ref} \left( z \right) = L_{flat} \left( z \right) \left\{ 1+ L_{flat} \left( z \right) \right\}^{-1}$, where $L_{flat} \left( z \right) = \mathrm{c2d} \circ \mathrm{f2i} \left(L_{Bode} \left( s; \psi \right)\right)$ for 
$\psi 
=
\begin{bmatrix}
    \phi_{m} & \omega_{c} \\
\end{bmatrix}$.
We use the particle swarm optimization to solve \eqref{Eq_minimize_J}, which is implemented in the MATLAB Global Optimization Toolbox (\texttt{particleswarm}). Note that the plant model is not used at all for the proposed controller design approach in all examples.

\begin{remark} \label{Remark_PSO}
    The PSO algorithm is used in this study because the nature of the PSO algorithm is well-suited to the optimization problem \eqref{Eq_minimize_J}. In particular, the problem \eqref{Eq_minimize_J} is non-convex, which requires a careful search over the solution space. Thus, population-based optimization strategies, including PSO, are well-suited for addressing \eqref{Eq_minimize_J}. Notably, we can easily use population-based algorithms as a solver for \eqref{Eq_minimize_J}, as the value of $J \left(\theta\right)$ can be evaluated without conducting closed-loop experiments with the controllers under test. This feature greatly reduces the cost to evaluate the objective function value compared with conventional optimization-based approaches that must evaluate it by actually implementing each controller to be tested into the real-world system and conducting closed-loop control experiments. Here, PSO is such a common population-based algorithm that it has already been implemented in commercial numerical software. Hence, using PSO as the solver in the verification examples shows the high practicality of the proposed approach. 
\end{remark}\label{S_Numerical_example}

\subsection{Example 1: High-order process (FO control vs. IO control)}\label{S_Example-1}
\begin{table}[]
    \centering
    \caption{Conditions for Example 1} \label{Table_Condition_Ex1}
    \begin{tabular}{c|c}
        \hline
        \multirow{2}{*}{$\theta_{0}$}  &  
        $\begin{bmatrix}
            K_{p} & K_{i} & K_{d} \\
        \end{bmatrix} =
        \begin{bmatrix}
            1 & 0 & 0 \\
        \end{bmatrix}$ \\
                                & 
        $\begin{bmatrix}
            K_{fp} & K_{fi} & \lambda & K_{fd} & \mu \\
        \end{bmatrix} =
        \begin{bmatrix}
            1 & 0 & 1 & 0 & 1\\
        \end{bmatrix}$  \\  \hline
        \multirow{2}{*}{Search range} & 
        $K_{p},K_{i},K_{d} \in 
        \begin{bmatrix}
            0 & 5 \\
        \end{bmatrix}$  \\
        & $K_{fp},K_{fi},K_{fd} \in 
        \begin{bmatrix}
            0 & 5 \\
        \end{bmatrix}$, \;
        $\lambda,\mu \in 
        \begin{bmatrix}
            0 & 2 \\
        \end{bmatrix}$  \\ \hline
        $J\left(\theta_{0}\right)$ & 
        $9.8820 \times 10^{2}$ \\ \hline
    \end{tabular}
\end{table}

The controlled plant $P\left(z\right)$ is
\begin{equation} \label{Eq_P_z_3rd}
    P \left( z \right) = \mathrm{c2d} \left( P_{3rd} \left(s\right) \right)
\end{equation}
where $P_{3rd}\left(s\right)$ is the following high-order process model \cite{YGE2019}:
\begin{equation}
    \label{eq19}
    P_{3rd} \left(s\right) = \frac{9}{\left(s+1\right)\left(s^{2} + 2s + 9\right)}.
\end{equation}
The parameters of $L_{Bode} \left( s; \psi \right)$ to design $M_{ref} \left( z \right)$ are set as 
$\psi 
=
\begin{bmatrix}
    \phi_{m} & \omega_{c} \\
\end{bmatrix}
=
\begin{bmatrix}
    80\tcdegree & 1 \, \mathrm{rad/s} \\
\end{bmatrix}$.
The sampling time $t_{s}$ $\mathrm{[s]}$ is set as $t_{s}=10^{-2}$ for the discretization of the plant, the reference model, and the controller.

In this example, we demonstrate the superiority of FO control to IO control. Specifically, we compare the performance of the FO-PID controller 
$C_{FOPID} \left( z; \theta \right) = \mathrm{c2d} \circ \mathrm{f2i} \left( C_{FOPID}\left( s; \theta \right) \right)$
with that of the traditional IO-PID controller
$C_{IOPID} \left( z; \theta \right) = \mathrm{c2d} \left( C_{IOPID}\left( s; \theta \right) \right)$.
Here, $C_{IOPID}\left( s; \theta \right)$ and $C_{FOPID}\left( s; \theta \right)$ are as follows \cite{DVCH2019}:
\begin{equation} \label{Eq_IOPID}
    C_{IOPID}\left( s; \theta \right) = K_{p} + K_{i}\frac{1}{s} + K_{d}\frac{s}{1 + \tau_{i}s}
\end{equation}
\begin{equation} \label{Eq_FOPID}
    C_{FOPID}\left( s; \theta \right) = K_{fp} + K_{fi}\frac{1}{s^{\lambda}} + K_{fd}\frac{s^{\mu}}{1 + \tau_{f}s^{\mu}}
\end{equation}
where $K_{p}, K_{i}, K_{d}, K_{fp}, K_{fi}, K_{fd}, \lambda, \mu, \tau_{i}, \tau_{f} \in \mathbb{R}_{+}\cup \left\{0\right\}$. The FO-PID controller is a typical example of the FC-based control scheme \cite{XTZJ2021}. In this study, we set $\tau_{i} = \tau_{f} = t_{s}$. 

\begin{figure}
    \begin{center}
    \includegraphics[width=7.0cm]{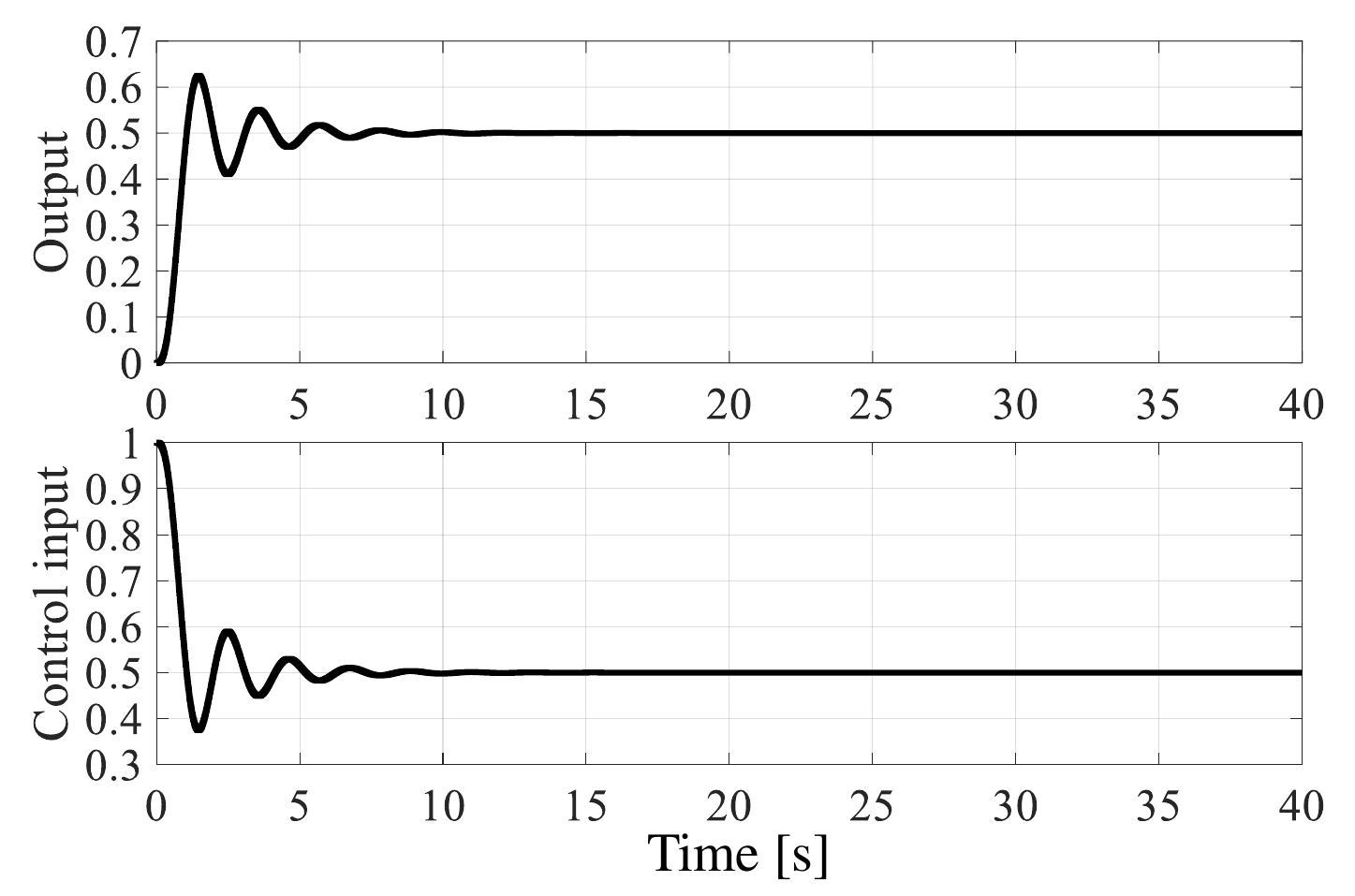}
    \caption{\protect{Initial input/output data for the proposed approach (Example 1)}}
    \label{Fig_Ini_data_Ex1}
    \end{center}
\end{figure}

The tuning conditions are listed in Table \ref{Table_Condition_Ex1}. The IO approximation is performed via the Oustaloup recursive filter (order: 5; valid frequency range $\left( \omega_{b}, \omega_{h} \right) = \left( 10^{-4}, 10^{4}\right) \; \mathrm{rad/s}$) for both $C_{FOPID} \left( z; \theta \right)$ and $M_{ref}\left(z\right)$. 
Fig. \ref{Fig_Ini_data_Ex1} shows the input/output data $u_{[0:N]}^{D}$ and $y_{[0:N]}^{D}$ for tuning both $C_{IOPID}\left(z; \theta\right)$ and $C_{FOPID}\left(z; \theta\right)$. The input/output data are collected through the closed-loop control simulation with the unit step as the reference input $r_{[0:N]}^{D}$ and the initial controller parameter $\theta_{0}$ shown in Table \ref{Table_Condition_Ex1}, where the simulation time is $40 \, \mathrm{s}$. Note that the initial data for the controller tuning are the same to each other in the FO-PID controller and the IO-PID controller, since 
$C_{IOPID}\left(z; \theta_{0}\right) = C_{FOPID}\left(z; \theta_{0}\right)$.

\begin{figure}
    \begin{center}
    \includegraphics[width=8.1cm]{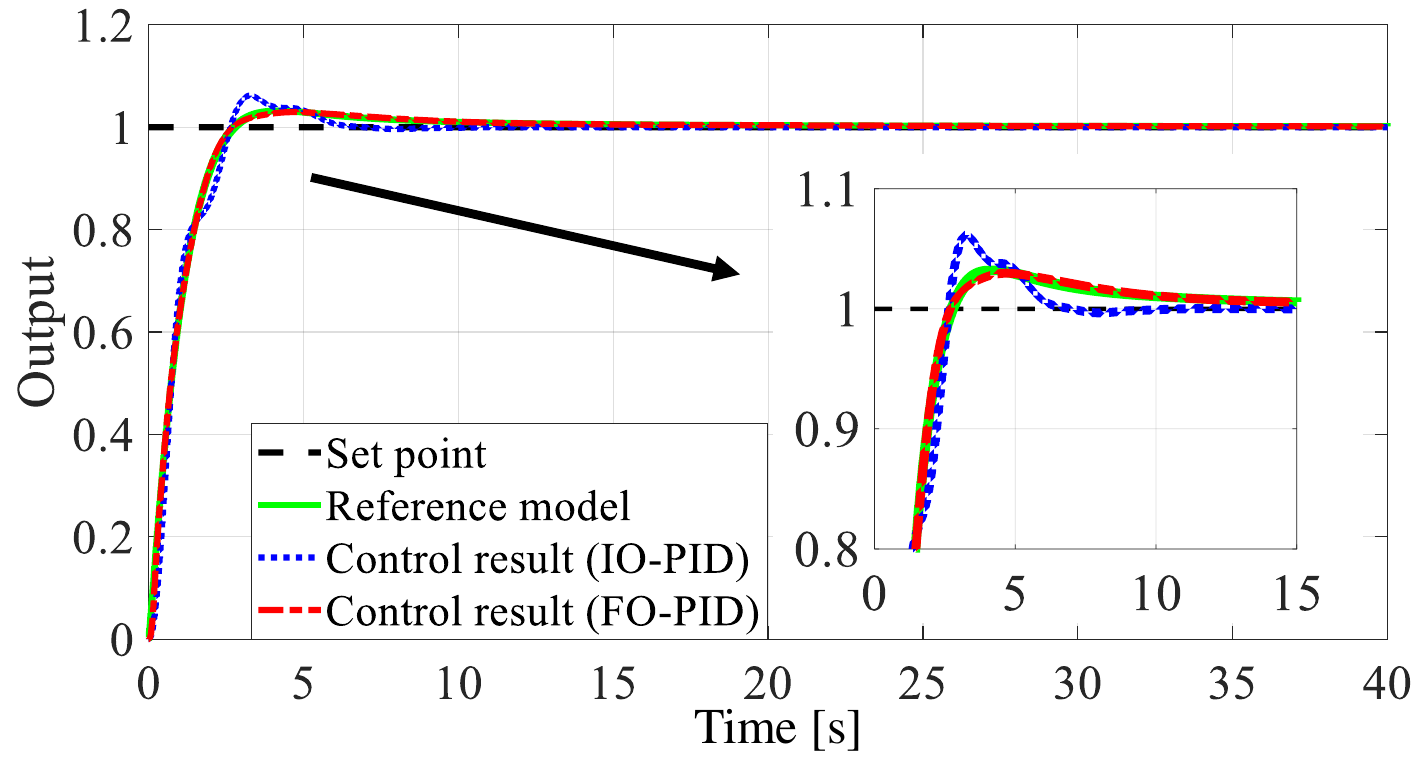}
    \caption{\protect{Control results due to the tuned controllers (Example 1). The green and red lines almost overlap.}}
    \label{Fig_Control_result_Ex1}
    \end{center}
\end{figure}

\begin{figure}
    \begin{center}
    \includegraphics[width=8.6cm]{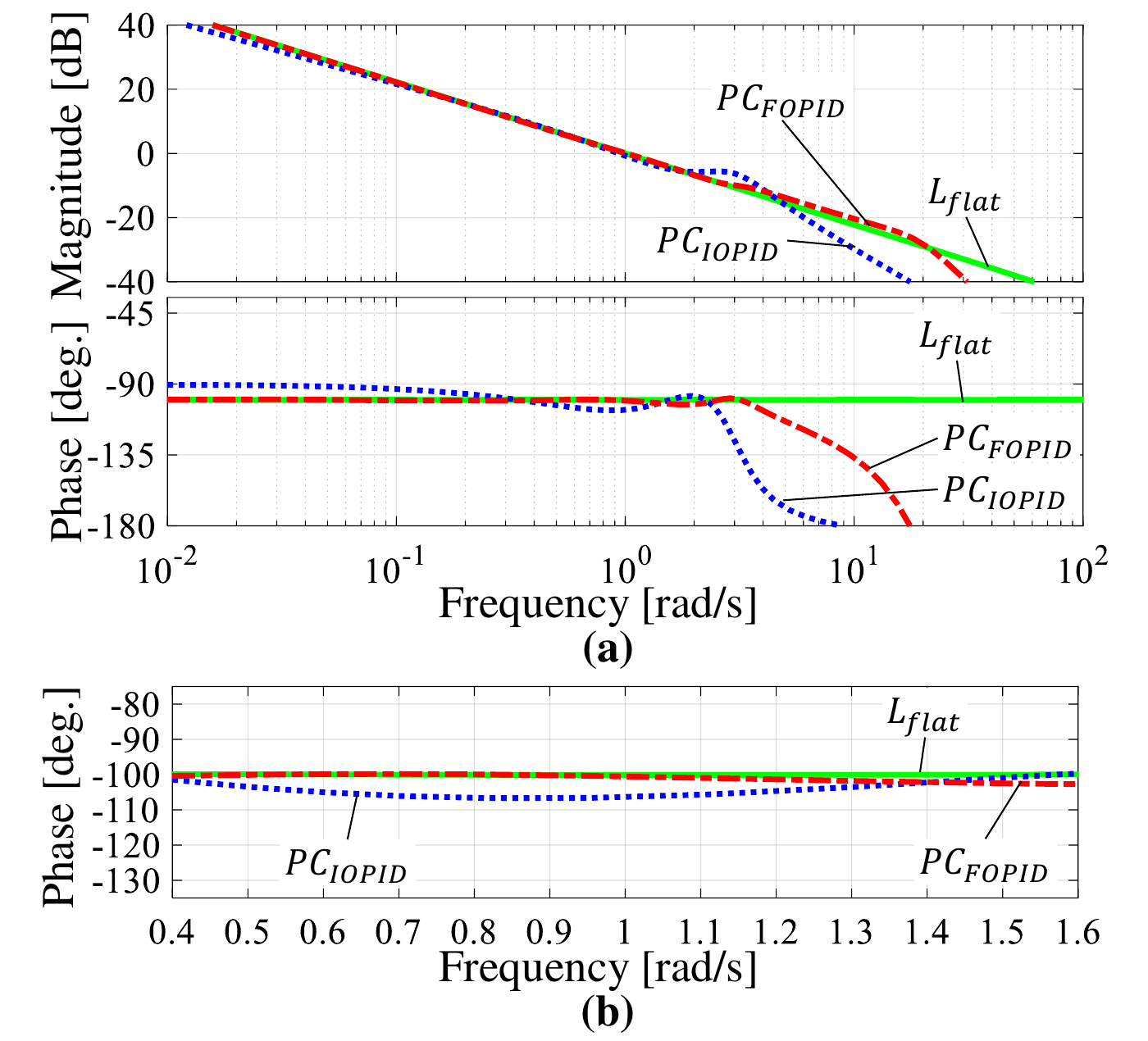}
    \caption{\protect{Bode plots of $L_{flat} \left(z\right)$, 
    $P\left(z\right)C_{IOPID}\left(z; \theta^{\ast}\right)$, and
    $P\left(z\right)C_{FOPID}\left(z; \theta^{\ast}\right)$ (Example 1): 
    (a) overview, and (b) enlarged phase plots around $1 \, \mathrm{rad/s}$.}}
    \label{Fig_Bode_Ex1}
    \end{center}
\end{figure}

Table \ref{Table_Tuning_result_Ex1} summarizes the tuning results for $C_{IOPID}\left(z; \theta\right)$ and $C_{FOPID}\left(z; \theta\right)$ provided by the proposed approach (i.e., Iso-IDFRIT). Table \ref{Table_Frequency_Ex1} lists the phase margin $\phi_{m}$ and the gain crossover frequency $\omega_{c}$ provided by the tuned controllers. Fig. \ref{Fig_Control_result_Ex1} demonstrates the control results obtained by the resultant controllers. The Bode plots of $L_{flat}\left(z\right)$, $P\left(z\right)C_{IOPID}\left(z; \theta^{\ast}\right)$, and $P\left(z\right)C_{FOPID}\left(z; \theta^{\ast}\right)$ are shown in Fig. \ref{Fig_Bode_Ex1}. In Fig. \ref{Fig_Control_result_Ex1}, the black line indicates the setpoint reference input. In Figs. \ref{Fig_Control_result_Ex1} and \ref{Fig_Bode_Ex1}, the green, blue, and red lines demonstrate the results due to the reference model ($M_{ref}\left(z\right)$ in Fig. \ref{Fig_Control_result_Ex1} and $L_{flat}\left(z\right)$ in Fig. \ref{Fig_Bode_Ex1}), $C_{IOPID}\left(z; \theta^{\ast}\right)$, and $C_{FOPID}\left(z; \theta^{\ast}\right)$, respectively. Tables \ref{Table_Tuning_result_Ex1} and \ref{Table_Frequency_Ex1} and Figs. \ref{Fig_Control_result_Ex1} and \ref{Fig_Bode_Ex1} clearly show that the proposed approach successfully provides the IO- and FO-PID controllers yielding the desired closed-loop property. Note that the closed-loop system due to the FO-PID controller achieves the closer specifications ($\phi_{m}$, $\omega_{c}$, the flat-phase) to the desired ones than that due to the IO-PID controller.

{

\tabcolsep = 4pt  

\begin{table}[]
    \centering
    \caption{Tuning results (Example 1)} \label{Table_Tuning_result_Ex1}
    \begin{tabular}{cc|c}
    \hline
    \multirow{2}{*}{$\theta^{\ast}$} & IO & 
    $\begin{bmatrix}
        8.0397 \times 10^{-1} & 1.2125 & 3.3528 \times 10^{-1} \\
    \end{bmatrix}$ \\
                                     & FO & 
    $\begin{bmatrix}
        1.3239 & 1.0370 & 1.1010 & 2.3253 \times 10^{-1} & 1.5465 \\
    \end{bmatrix}$ \\
    \multirow{2}{*}{$J \left(\theta^{\ast}\right)$}  & IO & 
    $1.0172$ \\
                                                     & FO & 
    $4.2744 \times 10^{-2}$ \\ \hline
    \end{tabular}
\end{table}
}  

\begin{table}[]
    \centering
    \caption{Frequency domain property provided by the tuned controllers (Example 1).} \label{Table_Frequency_Ex1}
    \begin{tabular}{c|c|c}
        \hline
                 & IO    & FO    \\ \hline
        Phase margin $\phi_{m} \, [\mathrm{deg.}]$ & 73.3411 & 79.4165 \\
        Gain crossover frequency $\omega_{c} \, [\mathrm{rad/s}]$   & 0.9371 & 1.0135 \\ \hline
    \end{tabular}
\end{table}  

\begin{figure}
    \begin{center}
    \includegraphics[width=7.4cm]{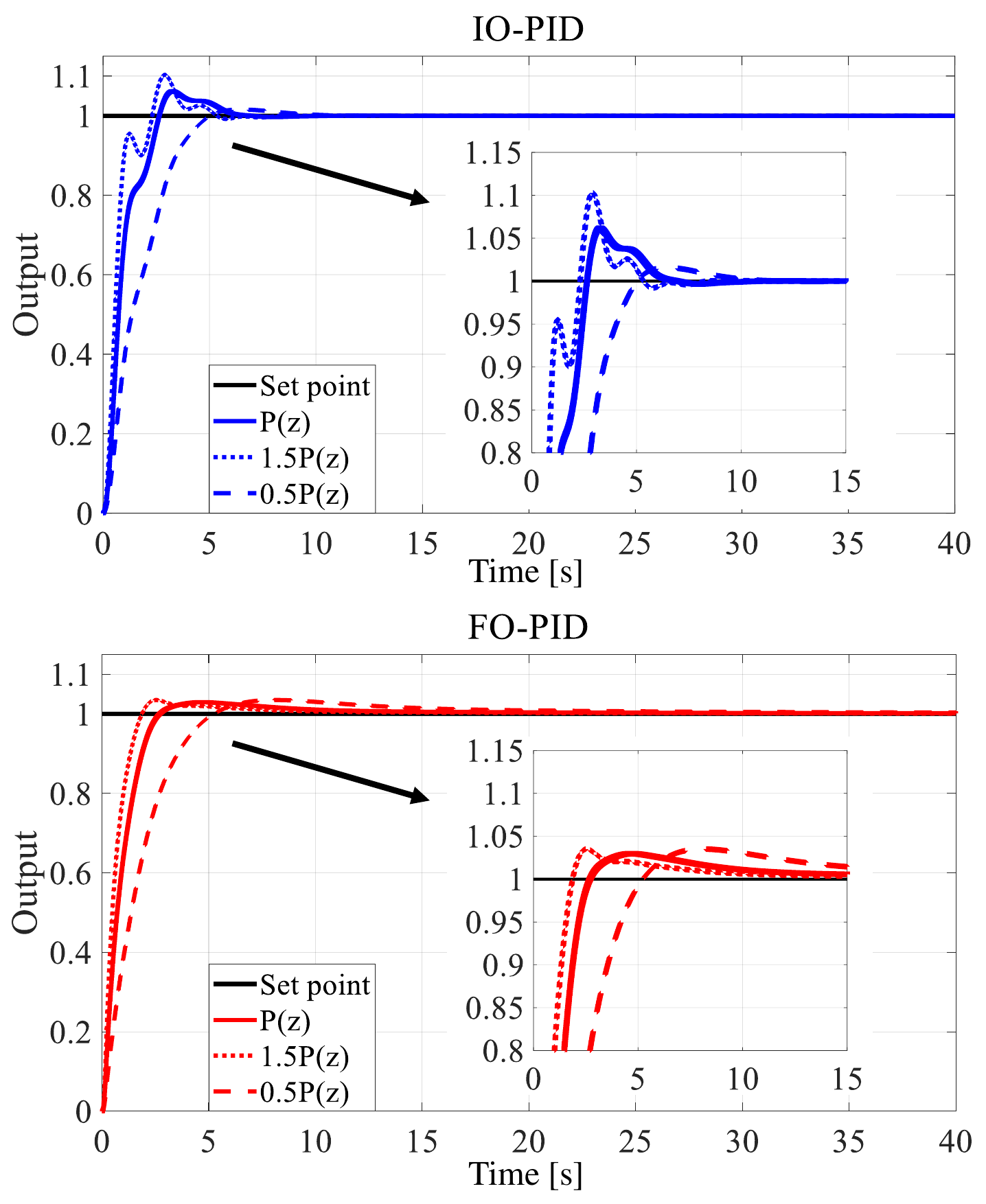}
    \caption{\protect{Robustness to the plant gain variation (Example 1).}}
    \label{Fig_Robustness_Ex1}
    \end{center}
\end{figure}

Fig. \ref{Fig_Robustness_Ex1} demonstrates the robustness to the plant gain variation of the controllers tuned by the proposed approach. The black line indicates the setpoint reference input. The solid lines, the dotted lines, and the dashed lines represent the control results for $P\left(z\right)$ (i.e., the nominal plant), $1.5P\left(z\right)$ (i.e., the plant with increased gain), and $0.5P\left(z\right)$ (i.e., the plant with decreased gain), respectively. The gain crossover frequencies provided by $C_{IOPID}\left(z; \theta^{\ast}\right)$ are $1.3095 \, \mathrm{rad/s}$ for $1.5P\left(z\right)$ and $5.3990 \, \mathrm{rad/s}$ for $0.5P\left(z\right)$, whereas these due to $C_{FOPID}\left(z; \theta^{\ast}\right)$ are $1.4618 \, \mathrm{rad/s}$ for $1.5P\left(z\right)$ and $5.2886 \, \mathrm{rad/s}$ for $0.5P\left(z\right)$. Fig. \ref{Fig_Robustness_Ex1} clearly shows that, although both $C_{IOPID}\left(z; \theta^{\ast}\right)$ and $C_{FOPID}\left(z; \theta^{\ast}\right)$ do not destabilize the closed-loop system, $C_{FOPID}\left(z; \theta^{\ast}\right)$ provides much stronger robustness of the overshoot amount to the plant gain variation, i.e., the iso-damping robustness, than $C_{IOPID}\left(z; \theta^{\ast}\right)$.

\subsection{Example 2: Soft robot (superiority over the conventional approaches)}\label{S_Example-2}
One representative virtue of the proposed approach is the freedom from the necessity of the plant model and the model reduction. This feature allows the proposed approach to achieve better control performance and robustness than the reduced-model-based controller design scheme. Note that almost all conventional analytical and graphical design methodologies rely on the reduced order model, as discussed in Section \ref{S_Related_Works}. We demonstrate the superiority and robustness of the proposed approach, compared to the traditional iso-damping controller design based on the reduced model. Throughout this example, the Oustaloup recursive filter is employed for the IO approximation (order: 7; valid frequency range 
$\left( \omega_{b}, \omega_{h} \right) = \left( 10^{-3}, 10^{6}\right) \; \mathrm{rad/s}$)
for both controllers and reference models. We discretize the plant, the reference model, and the controller in the sampling time $t_{s}=10^{-2} \, \mathrm{s}$.

\begin{figure}
    \begin{center}
    \includegraphics[width=7.0cm]{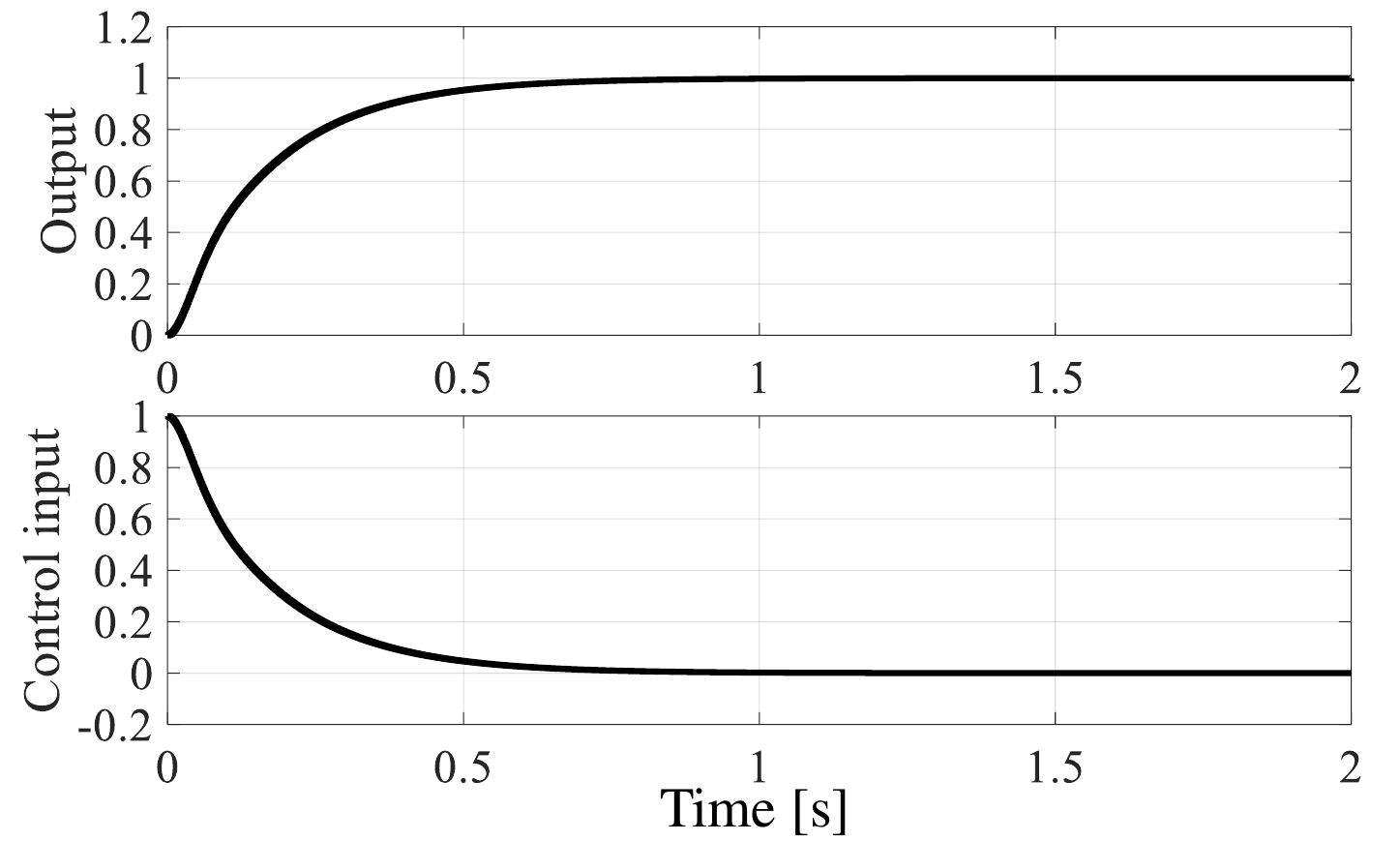}
    \caption{\protect{Initial input/output data for the proposed approach (Example 2).}}
    \label{Fig_Ini_data_Ex2}
    \end{center}
\end{figure}

The controlled plant $P\left(z\right)$ is 
\begin{equation} \label{Eq_P_z_full}
    P\left( z \right) = \mathrm{c2d}\left( P_{full}\left(s\right) \right)
\end{equation}
where $P_{full}\left(s\right)$ is as follows, which is the soft robot model (see \cite{MMNB2020}):
\begin{equation} \label{Eq_P_s_full}
    P_{full}\left( s \right) 
=
\frac{6}{s}\frac{54.893316s + 2048.6337}{s^{2} + 67.066887s + 2048.7922}.
\end{equation}
The parameters of $L_{Bode} \left( s; \psi \right)$ to design $M_{ref} \left( z \right)$ are set as 
$\psi 
=
\begin{bmatrix}
    \phi_{m} & \omega_{c} \\
\end{bmatrix}
=
\begin{bmatrix}
    60\tcdegree & 12 \, \mathrm{rad/s} \\
\end{bmatrix}$.
The controller class is chosen as the FO-PI controller 
$C_{FOPI} \left( z; \theta \right) = \mathrm{c2d} \circ \mathrm{f2i}\left( C_{FOPI}\left(s; \theta\right) \right)$. 
Here,
\begin{equation} \label{Eq_FOPI}
    C_{FOPI}\left( s; \theta \right) 
=
K_{fp} + K_{fi} \frac{1}{s^{\lambda}}
\end{equation}
where $K_{fp}, K_{fi}, \lambda \in \mathbb{R}_{+}\cup \left\{0\right\}$. The tuning condition is listed in Table \ref{Table_Condition_Ex2}. Fig. \ref{Fig_Ini_data_Ex2} shows the initial input/output data used for the proposed approach. These were collected through the closed-loop control simulation with $C_{FOPI} \left( z; \theta_{0} \right)$ as the initial controller and the unit step as the reference input $r_{[0:N]}^{D}$, where the simulation time was $2 \, \mathrm{s}$. Note that the initial data is collected on the basis of the actual plant $P\left(z\right)$ in \eqref{Eq_P_z_full}.

\begin{table}[]
    \centering
    \caption{Conditions for Example 2.} \label{Table_Condition_Ex2}
    \begin{tabular}{c|c}
        \hline
            $\theta_{0}$ & 
            $\begin{bmatrix}
             K_{fp} & K_{fi} & \lambda \\
             \end{bmatrix} =
             \begin{bmatrix}
              1 & 0 & 1 \\
             \end{bmatrix}$ \\
            Search range   & $K_{fp}, K_{fi} \in  
                \begin{bmatrix}
                0 & 15 \\
                \end{bmatrix} $,\; 
                $\lambda \in  
                \begin{bmatrix}
                0 & 2 \\
                \end{bmatrix} $\\
            $J \left( \theta_{0} \right)$ & $3.8646 \times 10^{1}$ \\ \hline
        \end{tabular}
\end{table}  

We compare the controller tuned by the proposed approach with that tuned by the \emph{iso-m} method as the representative graphical tuning approach for FO controllers \cite{MMNB2020}. Specifically, based on the iso-m method \cite{MMNB2020}, the controller $C_{FOPI}\left(s; \theta\right)$ is tuned using the following reduced plant model 
\begin{equation} \label{Eq_P_s_reduced}
    P_{reduced}\left( s \right) 
=
\frac{6}{s}\frac{54.89}{s + 54.89}
\end{equation}
to achieve
$\begin{bmatrix}
    \phi_{m} & \omega_{c} \\
\end{bmatrix}
=
\begin{bmatrix}
    60\tcdegree & 12 \, \mathrm{rad/s} \\
\end{bmatrix}$
(the details of the reduced model is described in \cite{MMNB2020}). As a result, the controller parameter is determined as 
$\theta_{iso-m} = 
\begin{bmatrix}
    K_{fp} & K_{fi} & \lambda \\
\end{bmatrix} =
\begin{bmatrix}
    1.76 & 4.7872 & 0.81 \\
\end{bmatrix}$ \cite{MMNB2020}. Note that the controller $C_{FOPI}\left( s ; \theta_{iso-m}\right)$ must be IO-approximated and discretized to implement in practice. Therefore, we compare the control performance and robustness of
$C_{proposed} \left(z\right) \coloneqq \mathrm{c2d} \circ \mathrm{f2i} \left( C_{FOPI}\left( s ; \theta^{\ast}\right) \right)$
with those of
$C_{iso-m} \left(z\right) \coloneqq \mathrm{c2d} \circ \mathrm{f2i} \left( C_{FOPI}\left( s ; \theta_{iso-m}\right) \right)$, 
where $\theta^{\ast}$ is the controller parameter designed by the proposed approach.

\begin{table}[]
    \centering
    \caption{Tuning result of the proposed approach (Example 2).} \label{Table_Tuning_result_Ex2}
    \begin{tabular}{c|c}
        \hline
        $\theta^{\ast}$ & 
        $\begin{bmatrix}
          8.8086 \times 10^{-1} & 3.8808 & 4.7498 \times 10^{-1} \\
         \end{bmatrix}$ \\
        $J\left(\theta^{\ast}\right)$  & $2.0231 \times 10^{-1}$ \\ \hline
        \end{tabular}
\end{table}  

\begin{table}[]
    \centering
    \caption{Frequency domain property with the actual plant $P \left( z \right)$ (Example 2).}
    \label{Table_Frequency_Ex2}
    \begin{tabular}{c|c|c}
        \hline
                 & $C_{iso-m}$    & $C_{proposed}$    \\ \hline
        Phase margin $\phi_{m} \, [\mathrm{deg.}]$ & 67.7359 & 60.1370 \\
        Gain crossover frequency $\omega_{c} \, [\mathrm{rad/s}]$   & 12.7008 & 12.0696 \\ \hline
        \end{tabular}
\end{table}  

\begin{figure}
    \begin{center}
    \includegraphics[width=8.0cm]{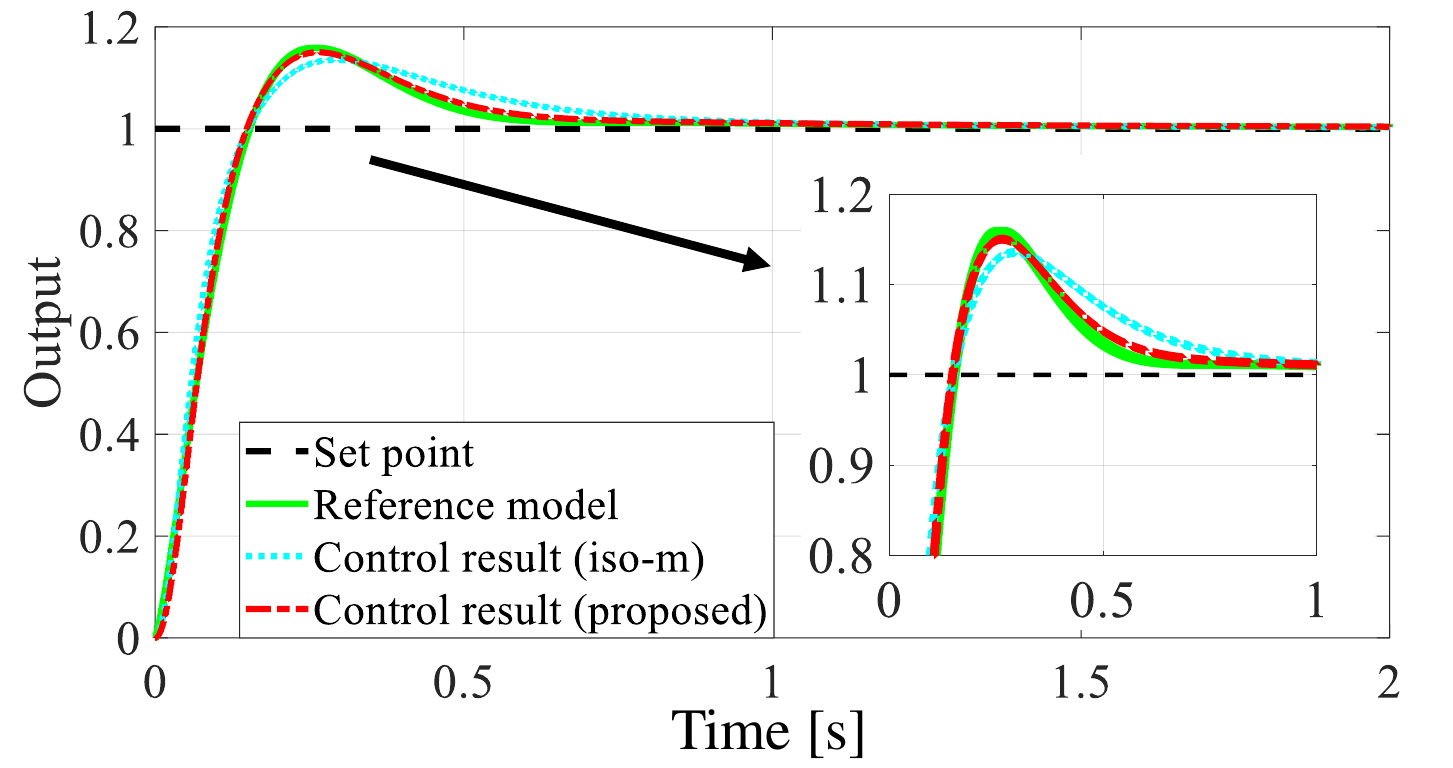}
    \caption{\protect{Control results due to $C_{iso-m} \left(z \right)$ and $C_{proposed} \left(z \right)$ (Example 2). The green and red lines almost overlap}}
    \label{Fig_Control_result_Ex2}
    \end{center}
\end{figure} 

\begin{figure}
    \begin{center}
    \includegraphics[width=7.9cm]{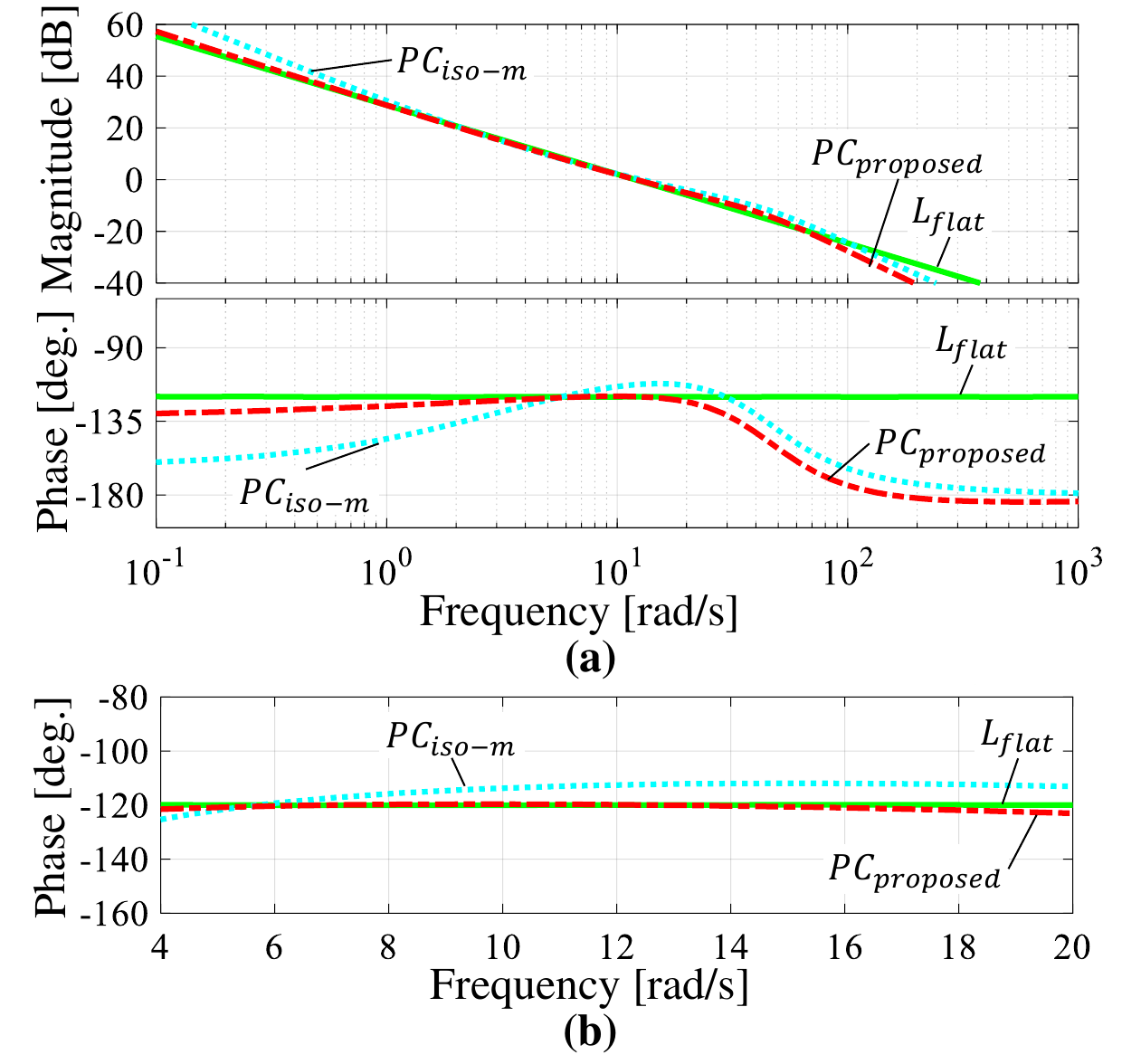}
    \caption{\protect{Bode plots of $L_{flat} \left(z\right)$, 
    $P\left(z\right)C_{iso-m}\left(z; \theta^{\ast}\right)$, and 
    $P\left(z\right)C_{proposed}\left(z; \theta^{\ast}\right)$ (Example 2): 
    (a) overview, and (b) enlarged phase plots around $12 \, \mathrm{rad/s}$.}}
    \label{Fig_Bode_Ex2}
    \end{center}
\end{figure} 

The tuning result due to the proposed approach is summarized in Table \ref{Table_Tuning_result_Ex2}. Table \ref{Table_Frequency_Ex2} compares the phase margin $\phi_{m}$ and the gain crossover frequency $\omega_{c}$ provided by $C_{iso-m} \left(z\right)$ and $C_{proposed} \left(z\right)$. Fig. \ref{Fig_Control_result_Ex2} demonstrates the control results provided by $C_{iso-m} \left(z\right)$ and $C_{proposed} \left(z\right)$; Fig. \ref{Fig_Bode_Ex2} shows the Bode plots of $L_{flat}\left(z\right)$, $P\left(z\right)C_{iso-m}\left(z\right)$, and $P\left(z\right)C_{proposed}\left(z\right)$.  The setpoint reference is indicated by the black line in Fig. \ref{Fig_Control_result_Ex2}. In Figs. \ref{Fig_Control_result_Ex2} and \ref{Fig_Bode_Ex2}, the results due to the reference model, $C_{iso-m} \left(z\right)$, and $C_{proposed} \left(z\right)$ are described by the green, cyan, and red lines, respectively. Note that the results in Table \ref{Table_Frequency_Ex2} and Figs. \ref{Fig_Control_result_Ex2} and \ref{Fig_Bode_Ex2} are computed on the basis of not the reduced order plant $\mathrm{c2d}\left(P_{reduced}\left(s\right)\right)$ but the actual controlled plant $P\left(z\right)$. Results in Table \ref{Table_Frequency_Ex2} and Figs. \ref{Fig_Control_result_Ex2} and \ref{Fig_Bode_Ex2} clearly demonstrate that $C_{proposed} \left(z\right)$ achieves the closed-loop properties ($\phi_{m}$, $\omega_{c}$, the flat-phase) much closer to the desired ones than $C_{iso-m} \left(z\right)$.

\begin{figure}
    \begin{center}
    \includegraphics[width=7.7cm]{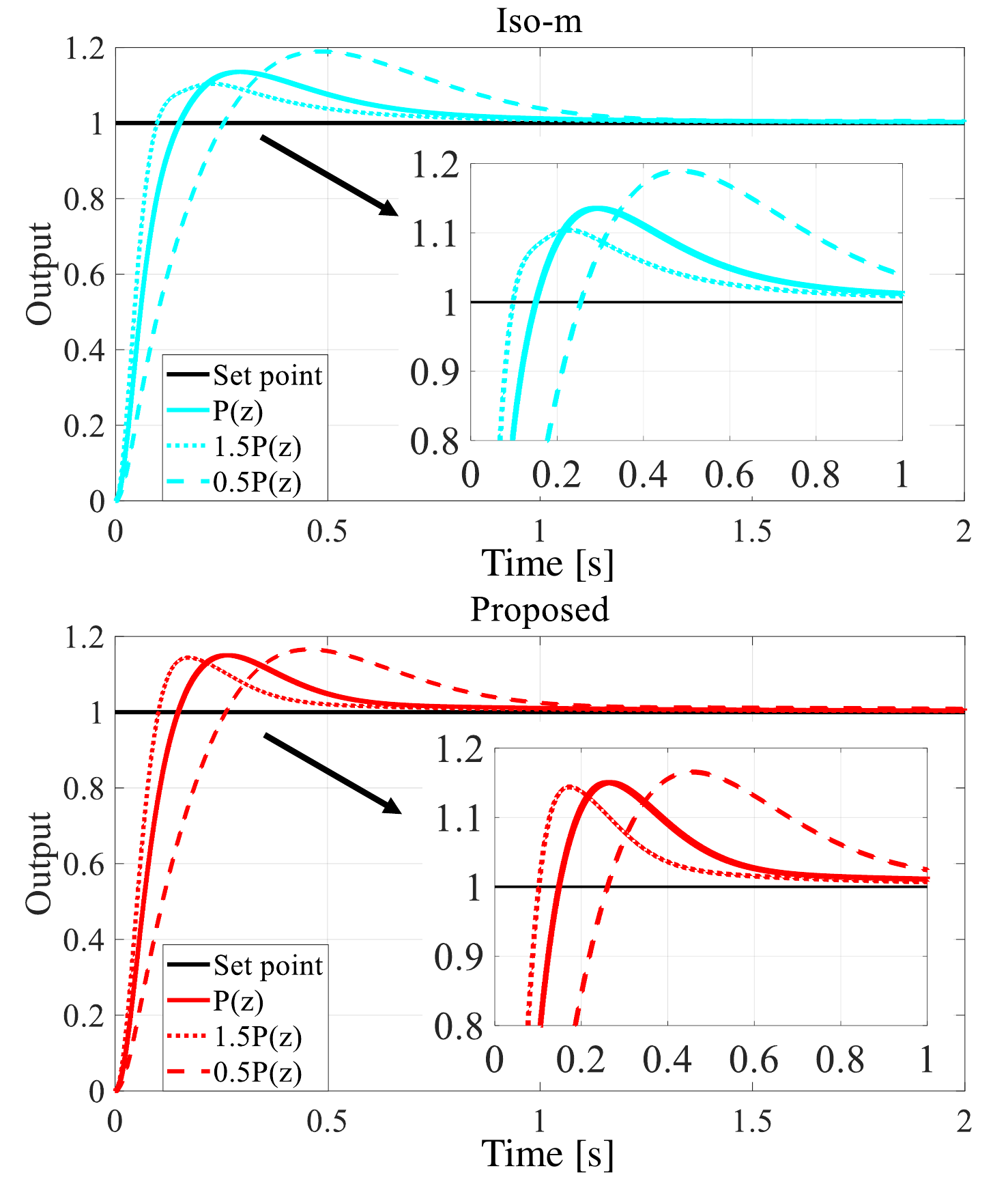}
    \caption{\protect{Robustness to the plant gain variation (Example 2).}}
    \label{Fig_Robustness_Ex2}
    \end{center}
\end{figure}

We also compare the iso-damping robustness of $C_{iso-m} \left(z\right)$ and $C_{proposed} \left(z\right)$. Fig. \ref{Fig_Robustness_Ex2} demonstrates the control results under the plant gain variation provided by $C_{iso-m} \left(z\right)$ and $C_{proposed} \left(z\right)$. Here, the black line indicates the setpoint reference input; the solid lines, the dotted lines, and the dashed lines represent the control results for $P\left(z\right)$ (i.e., the nominal plant), $1.5P\left(z\right)$ (i.e., the plant with increased gain), and $0.5P\left(z\right)$ (i.e., the plant with decreased gain), respectively. The gain crossover frequencies due to $C_{iso-m} \left(z\right)$ are $18.9349 \, \mathrm{rad/s}$ for $1.5P\left(z\right)$ and $6.8977 \, \mathrm{rad/s}$ for $0.5P\left(z\right)$; these provided by $C_{proposed} \left(z\right)$ are $17.0117 \, \mathrm{rad/s}$ for $1.5P\left(z\right)$ and $6.9209 \, \mathrm{rad/s}$ for $0.5P\left(z\right)$. As expected by the Bode plots in Fig. \ref{Fig_Bode_Ex2}, the characteristic of the closed-loop system due to $C_{proposed} \left(z\right)$ is closer to the ideal iso-damping property than that due to $C_{iso-m} \left(z\right)$.

To further highlight its advantage, the proposed approach is compared with IFT as a traditional data-driven FO controller tuning technique. Numerous studies have reported the effectiveness of the IFT approach for tuning FO controllers, e.g., \cite{MSK2019}. In this example, IFT seeks 
$\theta = \begin{bmatrix} K_{fp} & K_{fi} & \lambda \\ \end{bmatrix}$
that minimizes 
$E \left( \theta \right) = \left\lVert T \left( z; \theta^{\ast} \right)r_{[0:N]}^{D} - M_{ref}\left( z \right) r_{[0:N]}^{D} \right\rVert_{2}^{2} $, which is the same MR control problem as the proposed approach. Specifically, the IFT approach finds the minimizer of $E \left( \theta \right)$ by iteratively conducting closed-loop tests and updating $\theta$ via PSO; PSO is used to ensure fair comparison with the proposed approach. As a result, the IFT method provided
$\theta = \begin{bmatrix} 8.8086 \times 10^{-1} & 3.8808 & 4.7498 \times 10^{-1} \\ \end{bmatrix}$
, which is the same as that obtained by the proposed approach. 
This result stems from the fact that the proposed objective function \eqref{Eq_Loss_iso-IDFRIT} is an exact reformulation of $E \left( \theta \right)$ based on a single set of input and output data, as proved in Theorem \ref{Theorem_iso-IDFRIT}. This fact indicates that the proposed approach has a significant advantage over IFT for tuning FO controllers in that the required number of the closed-loop tests is reduced.

\subsection{Example 3: Experimental verification}\label{S_Example-3}
\begin{figure}
    \begin{center}
    \includegraphics[width=8.0cm]{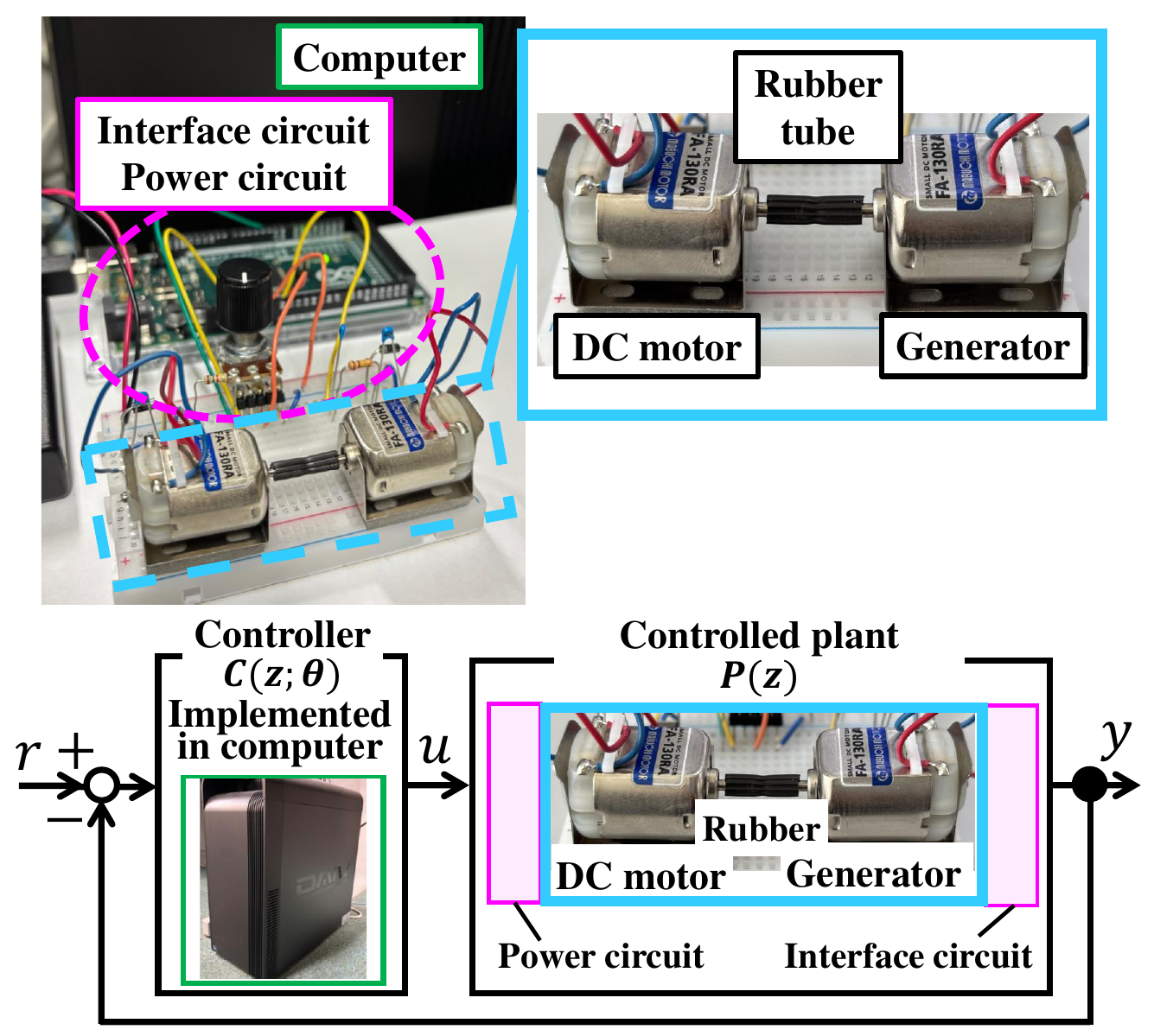}
    \caption{\protect{
            Experimental setup of Example 3. The control input $u$ is the command value of the voltage applied to drive the motor. The output $y$ is the voltage generated by the generator, which is proportional to the rotational speed of the generator.
    }}
    \label{Fig_Setup_Ex3}
    \end{center}
\end{figure}

We experimentally demonstrate the effectiveness of the proposed approach for real-world control systems. Figure \ref{Fig_Setup_Ex3} shows the experimental setup. The controlled plant consists of a motor and a generator, which are connected by a rubber tube. Both the motor and the generator are direct-current motors (Mabuchi Motor Co., Ltd. FA-130-RA-2270). In this experiment, the control input $u$ is the command value of the voltage applied to drive the motor, which is calculated by the controller. The output $y$ is the voltage generated by the generator, which is proportional to the rotational speed of the generator. The parameters of $L_{Bode} \left( s; \psi \right)$ to design $M_{ref} \left( z \right)$ are set as 
$\psi 
=
\begin{bmatrix}
\phi_{m} & \omega_{c} \\
\end{bmatrix}
=
\begin{bmatrix}
70\tcdegree & 3 \, \mathrm{rad/s} \\
\end{bmatrix}$.
The controller to be tuned is the FO-PI controller 
$C_{FOPI} \left( z; \theta \right) = \mathrm{c2d} \circ \mathrm{f2i}\left( C_{FOPI}\left(s; \theta\right) \right)$
, where $C_{FOPI}\left(s; \theta\right)$ is shown in \eqref{Eq_FOPI}. Here,
$\theta = \begin{bmatrix} K_{fp} & K_{fi} & \lambda \\ \end{bmatrix}$.
The Oustaloup recursive filter is used for the IO approximation order (order: 7; valid frequency range 
$\left( \omega_{b}, \omega_{h} \right) = \left( 10^{-5}, 10^{4}\right) \; \mathrm{rad/s}$) for both the controller and reference model. The sampling time $t_{s}$ is set to $10^{-2} \;\mathrm{s}$. 

\begin{figure}
    \begin{center}
    \includegraphics[width=7.2cm]{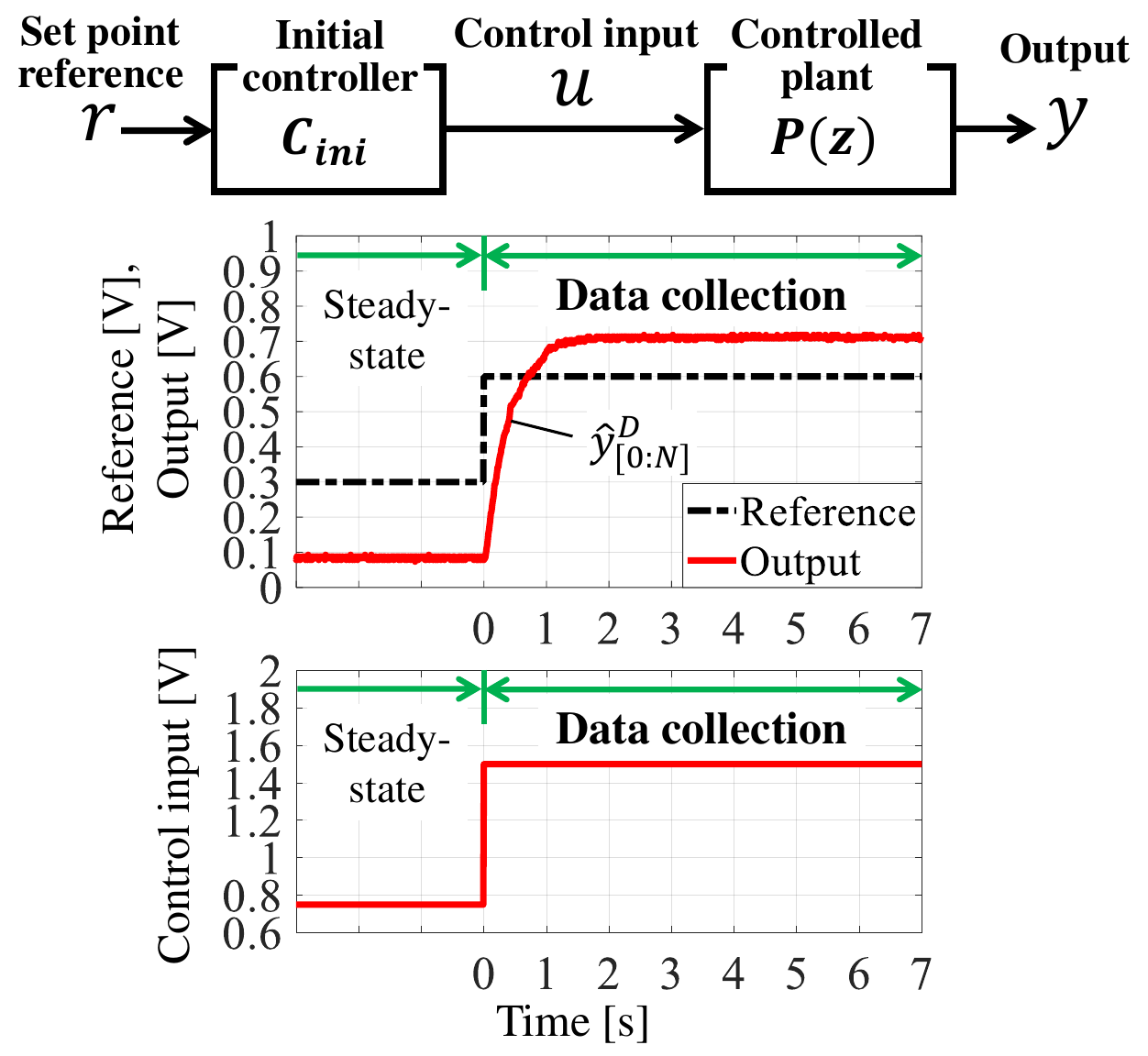}
    \caption{\protect{
            Sketch of the initial data collection procedure. The data is collected through an open-loop control experiment using the proportional controller $C_{ini} = 2.5$. The data collection begins when a step reference is applied to the controlled plant in a steady-state (i.e., the motor is rotating steadily).
    }}
    \label{Fig_DataCollection_Ex3}
    \end{center}
\end{figure}

\begin{figure}
    \begin{center}
    \includegraphics[width=6.8cm]{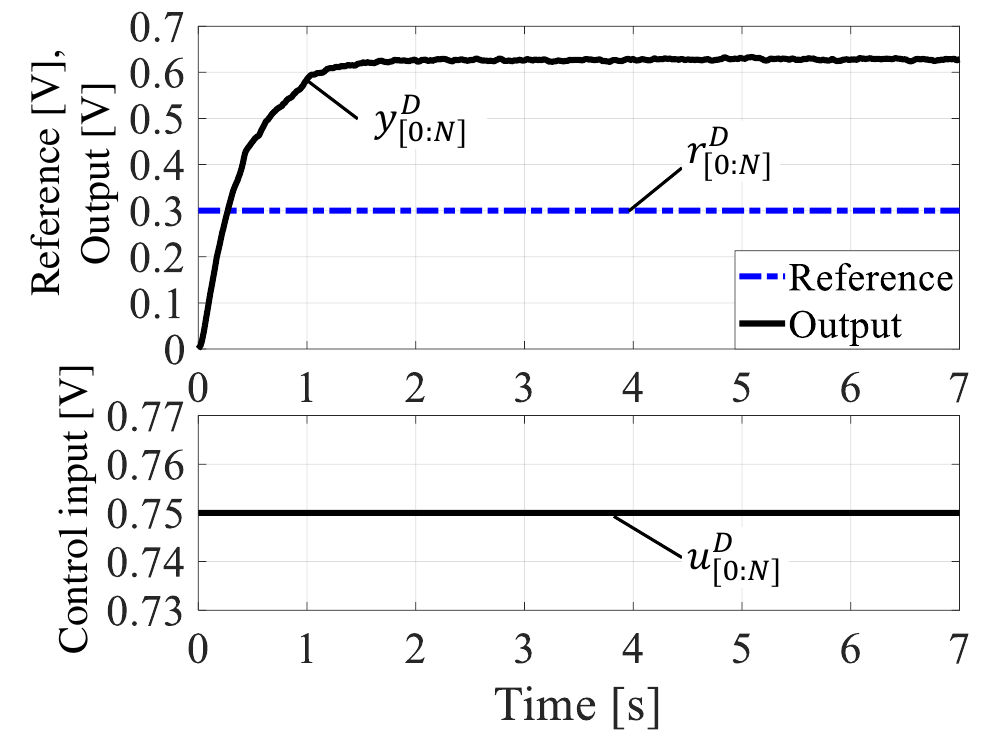}
    \caption{\protect{
            Input/output data $u_{[0:N]}^{D}$ and $y_{[0:N]}^{D}$ and setpoint reference $r_{[0:N]}^{D}$ for proposed controller tuning. The output data is denoised using the $L_{2}$ total variation denoising.
    }}
    \label{Fig_Ini_Data_Ex3}
    \end{center}
\end{figure}

In this example, data for the proposed approach is collected through an open-loop control experiment using the proportional controller $C_{ini} = 2.5$, as shown in Fig. \ref{Fig_DataCollection_Ex3}. The data collection begins when a step reference is applied to the controlled plant in a steady-state i.e., the motor is rotating steadily. As the output data is corrupted by measurement noise in real-world systems, the output data is mitigated using the $L_{2}$ total variation denoising technique \cite{YK2021_MC}:
\begin{equation} \label{Eq_Denoising}
    \mathrm{vec} \left( y_{[0:N]}^{D} \right) 
=
\left(I + \nu \chi^\top \chi \right)^{-1} \mathrm{vec} \left( \hat{y}_{[0:N]}^{D} \right) 

\end{equation}
\begin{equation} \label{Eq_Chi} 
    \chi \coloneqq
\begin{bmatrix}
    -1 & 1 & 0 & \hdots & 0 \\
    0 & -1 & 1 & \ddots & 0 \\
    \vdots & \ddots & \ddots & \ddots & 0 \\
    0 & \hdots & 0 & -1 & 1 \\
    0 & \hdots & 0 & 0 & 0 \\
\end{bmatrix}
\end{equation}
where $\nu \in \mathbb{R}_{+}$ (we set $\nu$ to $2$ in this example), $I$ is the identity matrix with the compatible dimension, 
$\hat{y}_{[0:N]}^{D}$ is the noisy output data shown in Fig. \ref{Fig_DataCollection_Ex3}, and $y_{[0:N]}^{D}$ is the denoised output data. The $L_{2}$ total variation denoising technique is often employed to handle noisy data in the data-driven controller design framework, owing to its computational efficiency. The details of this denoising approach have been discussed in \cite{YK2021_MC}. Figure \ref{Fig_Ini_Data_Ex3} shows the input/output data $u_{[0:N]}^{D}$ and $y_{[0:N]}^{D}$ and the setpoint reference $r_{[0:N]}^{D}$ for the proposed approach. 

\begin{figure}
    \begin{center}
    \includegraphics[width=8.0cm]{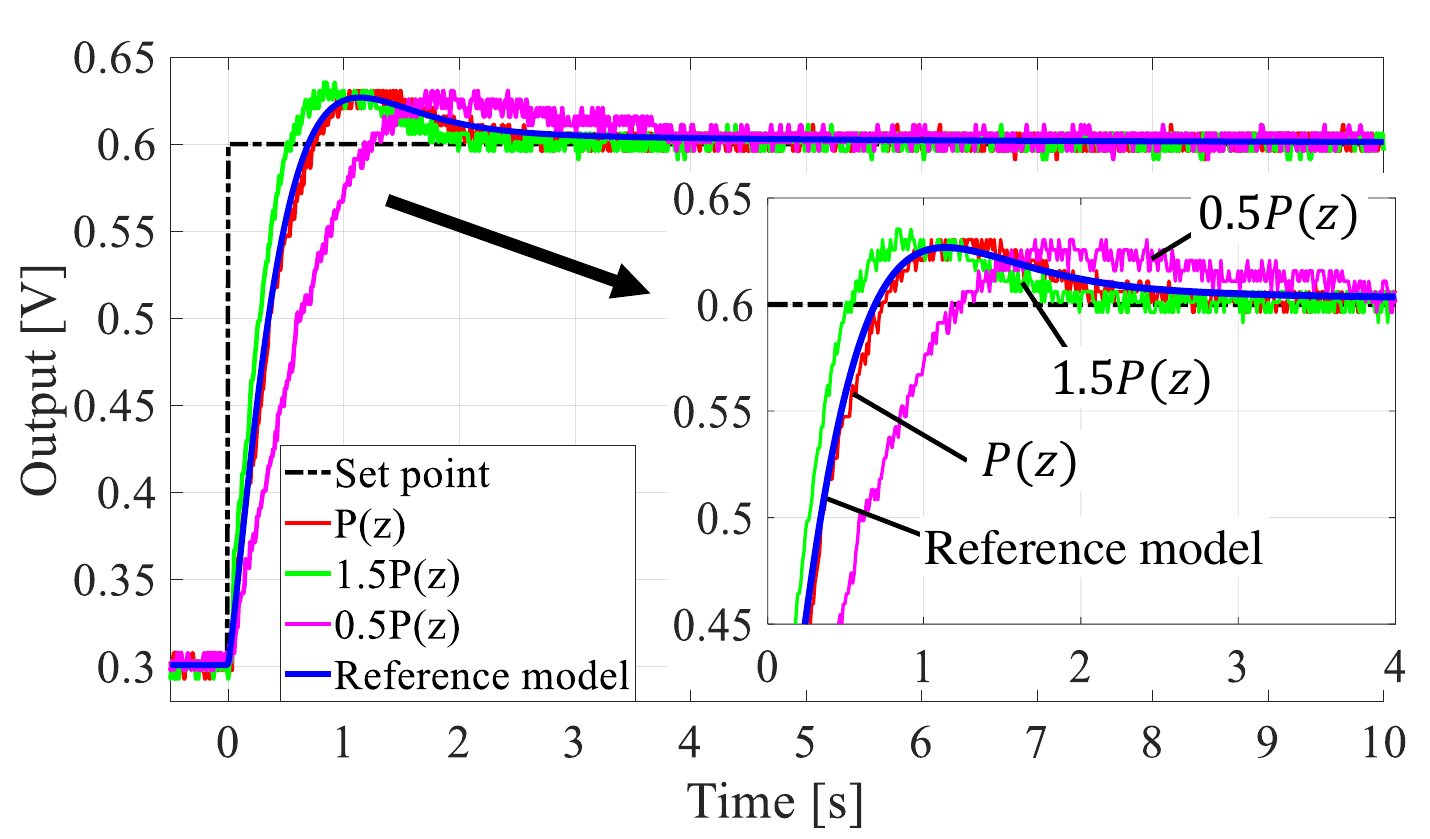}
    \caption{\protect{
            Control results due to the FO-PI controller tuned by the proposed approach. The comparison of the red, green, and magenta lines verifies the robustness of the overshoot amount against variations in plant gain. 
    }}
    \label{Fig_Control_result_Ex3}
    \end{center}
\end{figure}

The tuning result provided by the proposed approach is 
$\theta^{\ast} = \begin{bmatrix} K_{fp} & K_{fi} & \lambda \\ \end{bmatrix}
= \begin{bmatrix} 1.2026 & 4.9931 & 1.0911 \\ \end{bmatrix}$. 
Figure \ref{Fig_Control_result_Ex3} shows the control results obtained using $C_{FOPI} \left( z; \theta^{\ast} \right)$, i.e., the FO-PI controller tuned by the proposed approach. The black dashed-dotted line represents the reference input, which specifies the set point. The red, green, and magenta lines show the control results for $P\left(z\right)$ (nominal plant), $1.5P\left(z\right)$ (plant with a $50\%$ increase in gain), and $0.5P\left(z\right)$ (plant with a $50\%$ decrease in gain), respectively. The blue line represents the response of the reference model $M_{ref} \left( z \right)$. The comparison of the blue and red lines demonstrates that the proposed approach successfully yields the FO-PI controller that achieves the closed-loop response similar to the desired response given by $M_{ref} \left( z \right)$. As shown by the comparison of the red, green, and magenta lines, this tuning result achieves the iso-damping property. Thus, we validate the effectiveness of our proposed controller design approach for real-world control systems.

\section{Discussion}\label{S_Discussion}
In all examples, the proposed approach (Iso-IDFRIT) successfully provides controllers realizing the desired closed-loop characteristics specified by the designer-defined reference model. The controller achieves the iso-damping robustness. In the examples, no closed-loop destabilization is observed, showing that the closed-loop system is BIBO stable from a bounded reference input to a bounded output. Therefore, we confirm the validity of the proposed approach.

Unlike traditional analytical and graphical approaches, the proposed approach does not require plant models. Neither repeated experiments nor frequency response computations is necessary, whereas such procedures are indispensable for conventional data-driven and optimization-based approaches. In the present approach, the desired controller is automatically designed via solving the optimization problem defined based on only one-shot input/output data; the BIBO stability from a bounded reference input to a bounded output is explicitly considered for the IO-approximated and discretized (i.e., ready-to-implement) controllers. Hence, the proposed approach is a simple, practical, and reliable controller design technique to achieve the iso-damping property.

Example 1 shows not only the validity of the proposed approach but also the effectiveness of FO control for iso-damping robust control. Although both IO-PID and FO-PID controllers tuned by the present approach realize good control performance, the FO-PID controller outperforms the IO-PID controller from the viewpoint of the iso-damping robustness. This result is attributed to the fact that the reference model includes FO dynamics. Notably, the design burden of the FO-PID controller is not problematic compared with that of the IO-PID controller, owing to the simplicity of the proposed approach. The proposed approach contributes to the practical utilization of the virtues of FO control via reducing the implementation burden of FO controllers. 

In Example 2, the FO-PI controller tuned by the present approach outperforms that tuned on the basis of the reduced order model. In other words, the proposed approach achieves better optimality than the model-based approach. This result stems from the gap between the actual plant and the model for controller design. In general, practical systems include high-order dynamics. Neglecting the high-order dynamics and designing the controller on the basis of the reduced-order model may adversely affect the control performance, robustness, and stability. On the other hand, the present approach can avoid such issues related to the unmodeled dynamics, since it relies on not the plant model but the experimentally collected data.

Example 3 experimentally verifies the effectiveness of the proposed controller design approach in real-world systems. Real-world systems, including the one in this example, pose various challenges. For example, noise corrupting the measurement data may affect the data-based controller design approaches. In practice, the proposed approach is applicable to real-world systems by mitigating the noise in the data for controller design using well-established denoising techniques. In our experimental verification, the $L_{2}$ total variation denoising \cite{YK2021_MC} is adopted to mitigate the measurement noise. Note that this denoising technique is so simple and easy-to-use that incorporating this denoising procedure does not unduly complicate the proposed approach. Regarding issues related to the implementation of the proposed approach, mathematically proving BIBO stability is difficult in practical real-world systems due to unmodeled uncertainties and unexpected noise. Thus, in practice, it is important to reasonably infer the closed-loop properties before implementing the controller. From this perspective, the proposed approach yields a reliable controller, as we can reasonably infer that a controller parameter $\theta$ that minimizes $J\left(\theta\right)$ will yield $T\left(z; \theta\right)$ such that it is BIBO stable (see Remark \ref{Remark_BIBO_of_T}). In fact, in Example 3, the FOPI controller tuned by the proposed approach achieves good control performance and robustness without causing instability. Consequently, the proposed approach is a simple and effective technique for designing an iso-damping robust controller for real-world control systems. We will mathematically evaluate the effects of various uncertainties and constraints in real-world systems on the proposed approach in the future.

In this study, we assume that the reference model is given by the designer. That is, the designer must give the appropriate reference model. An important future direction may be the simplification of the design procedure for the reference model. For example, the flat-phase property is a frequency-domain specification, whereas it is more straightforward and intuitive to design the reference model in the time-domain. The Auto-DDC approach \cite{BF2023} may be effective for addressing this issue. One study has proposed the data-driven tuning of parameters of both the controller and the reference model \cite{YK2024_Access}. In the future, we will examine the Auto-DDC strategy and the simultaneous tuning technique of the controller and the reference model in iso-IDFRIT.

\section{Conclusion}\label{S_Conclusion}
This study has presented novel controller design strategy to achieve iso-damping robust control. The proposed approach is simpler, more practical, and reliable than conventional iso-damping controller design methodologies. The proposed approach designs the iso-damping controller on the basis of the MR-D3C framework, requiring only one-shot input/output data. The plant model, the computation of the frequency response, or repeated experiments are unnecessary. The present controller tuning scheme explicitly evaluates the BIBO stability of the resultant closed-loop system from a reference input to an output. The numerical and experimental studies have demonstrated the effectiveness of the proposed approach. Moreover, this study has verified the advantage of FO control for iso-damping robust control.

In the future, we will analyze the stability and performance of the present approach with a finite data length under measurement noise.

\bibliographystyle{IEEEtran}        
\bibliography{References_IEEETSMC}           

\vfill

\end{document}